\documentclass[prb,aps,twocolumn,epsfig,eqsecnum,floatfix]{revtex4}
\usepackage{epsfig}
\usepackage{times}
\usepackage{amsfonts}
\usepackage{amsmath}
\usepackage{graphicx}
\usepackage{subfigure}
\usepackage{wasysym}
\usepackage{bm}
\usepackage{pstricks}
\newlength{\myL}

\newcommand{\beq}{\begin{equation}}
\newcommand{\eeq}{\end{equation}}
\newcommand{\bea}{\begin{eqnarray}}
\newcommand{\eea}{\end{eqnarray}}

\begin{document}
\bibliographystyle{apsrev}

\title{Biot-Savart correlations in layered superconductors.}

\author{K. S. Raman$^1$, V. Oganesyan$^{2}$, S. L. Sondhi$^3$}

\affiliation{$^1$Department of Physics and Astronomy, University of California at Riverside, Riverside, CA 92507}

\affiliation{$^2$Department of Engineering Science and Physics, College of Staten Island, City University of New York, Staten Island, NY 10314}

\affiliation{$^{3}$Department of Physics, Princeton University, Princeton, NJ 08544}

\date{\today}

\begin{abstract}
    We discuss the superconductor to normal phase transition in an 
    infinite--layered, type--II superconductor in the limit where the Josephson 
    coupling between layers is negligible.  We model each layer as a
    neutral gas of thermally--excited pancake vortices and assume the dominant interlayer
    coupling is the electromagnetic interaction between the screening currents
    induced by these vortices.  Our main result, obtained by exactly solving the 
    leading order renormalization group flow, is that the phase transition in this 
    model is a Kosterlitz--Thouless transition despite being a three--dimensional system.
    While the transition itself is driven by the unbinding of two--dimensional 
    pancake vortices, an RG analysis of the low temperature
    phase and a mean--field theory of the high temperature phase reveal that
    both phases possess three--dimensional correlations.  An experimental
    consequence is that the jump in the measured in--plane superfluid
    stiffness, a universal quantity in 2d Kosterlitz--Thouless theory, 
    receives a small non--universal correction (of order 1\% in  
    Bi$_2$Sr$_2$CaCu$_2$O$_{8+x}$).  This overall picture places some claims 
    expressed in the literature on a more secure analytical footing and 
    resolves some conflicting views.    
\end{abstract}


\maketitle

\section{Introduction}
In this paper, we revisit the problem of the superconductor--normal phase 
transition of a layered, type--II superconductor in the limit where Josephson 
coupling between layers is negligible.  We model this system as an infinite
stack of superconducting planes, each layer containing a neutral gas of
thermally--excited, two--dimensional pancake vortices.\cite{Clem91}  We take
the viewpoint that the dominant mechanism coupling the layers is the 
long--range electromagnetic interaction\cite{Efetov79} between the screening currents
induced by these vortices.  We expect this model to be relevant to layered 
superconductors where the dominant mechanism by which the superconductivity
is lost (as the temperature is raised) is the loss of long--range order in the phase
of the order parameter.\cite{Emery02}  Candidate materials include the underdoped 
high--$T_c$ cuprates as well as layered structures made from conventional 
type--II superconductors.

One motivation for considering this model is that investigations of the 
superconductor--normal phase transition in the cuprates have revealed
3dXY critical exponents, a hallmark of Josephson coupling between the
planes, in only one case:  optimally--doped YBa$_2$Cu$_3$O$_{6+x}$, the least anisotropic of
these materials.\cite{Kamal94}  In underdoped Bi$_2$Sr$_2$CaCu$_2$O$_{8+x}$, 
the most anisotropic of these compounds, signatures suggestive of a Kosterlitz--Thouless (KT) transition have been seen\cite{Corson99, Osborn01, Li05} while recent measurements\cite{Broun07} 
on underdoped YBa$_2$Cu$_3$O$_{6+x}$, which is substantially more anisotropic
than the optimally--doped material though less so than underdoped 
Bi$_2$Sr$_2$CaCu$_2$O$_{8+x}$, show a
transition that is neither KT nor 3dXY nor any obvious interpolation in between.
\footnote{However, a jump in the superfluid stiffness characteristic of a KT 
transition has been seen in very thin films of underdoped YBa$_2$Cu$_3$O$_{6+x}$.\cite{Hetel07}}
While these observations do not conclusively show that Josephson coupling can be 
neglected\cite{Oganesyan06, Benfatto07}, they do suggest that investigations of different 
mechanisms for coupling the layers might be a fruitful line of attack if there is 
reason to suspect that Josephson coupling is very small.  The Biot--Savart interaction 
between screening currents in different layers is a long--range, three--dimensional 
coupling which is always present though its influence is usually 
assumed to be small compared to the Josephson term.    

A second motivation is a recent experiment\cite{Li07} on La$_{2-x}$Ba$_{x}$CuO$_4$ 
near its stripe-ordered state at $x=1/8$ where a 2d superconducting phase has been 
observed below an apparent Kosterlitz--Thouless transition temperature.  One 
proposal\cite{Berg07} suggests that under the right circumstances, the superconducting
state can occur with a finite wave vector, where the periodicity is in the same direction
as the charge order but with double the period.  Since the stripes in adjacent layers 
orient with a relative angle of 90$^{\circ}$, an orthogonality argument implies that the
Josephson coupling between first, second, and third neighbor planes is cancelled exactly
and further estimates\cite{Berg07} imply the residual terms are extremely small.
In such a scenario, we expect our model to be directly applicable to the 
experiments.\footnote{More precisely, since the
rotational symmetry of the plane is broken, we would expect the pancakes to be anisotropic 
in shape, since the coherence length will be different whether one is parallel or perpendicular
to the stripes.  Also, the interactions themselves might be anisotropic.  However, we expect
the leading long--distance behavior to be at least qualitatively captured by an isotropic model.}  
In a similar vein, there are a series of somewhat older experiments\cite{Safar92, Lopez94} on various 
cuprates in a dc flux transformer geometry where the observation of effectively
2d vortices was explained as the cutting of 3d vortex lines.  Since Josephson 
coupling is the strongest reason why pancake vortices tend to form 3d stacks/lines 
(as the penalty for breaking a line of pancakes bound by the Josephson 
interaction would be proportional to the system size $L$) one way to interpret the 
experiment is that the Josephson coupling is effectively ``cancelled" by the applied field.  
Once again, in such a case we would expect our model to apply.  

Our main result is that the superconductor to normal phase transition in a 
layered superconductor with an infinite number of electromagnetically--coupled 
layers is still a Kosterlitz--Thouless transition.  The mechanism resembles the
single--layer problem in that  (a) the transition occurs through the unbinding of 
two--dimensional pancake vortices and (b) the screening within an individual layer
at temperatures above $T_{KT}$  is not significantly different from an isolated 
two--dimensional system.  However, both the low and high temperature phases
have three--dimensional correlations and the jump in the in--plane superfluid
stiffness, as inferred from a penetration depth measurement, receives a small 
non--universal correction (of order 1$\%$).  Our results are obtained via a 
renormalization group study of the low temperature phase and a Debye--Huckel 
mean field theory of the high temperature state.  

The electromagnetic coupling may be formulated as an interaction between the
pancake vortices.\cite{Pearl64, Efetov79, Buzdin90, Artemenko90, Clem91, Fischer91, Brandt05}  The basic mechanism is that a vortex in one of the layers induces screening currents in the same and in other layers, which cause Biot--Savart forces on the other vortices.  For a single, literally two--dimensional layer, this vortex--vortex interaction is screened at distances larger than the magnetic
penetration depth $\lambda$.  For distances much less than $\lambda$, but greater than
the coherence length $\xi$, the interaction energy of two vortices of the same sign will
be repulsive and scale logarithmically with separation.  However, if the layer has a 
small but nonzero thickness $d$, the effective screening length becomes 
$\Lambda'=2\lambda^2/d \gg \lambda$.\cite{Pearl64, Brandt05}  $\Lambda'$ often 
exceeds the sample sizes considered in experiment and in such cases, the interaction 
is effectively logarithmic.  

For a layered system, the relevant screening length becomes $\Lambda=2\lambda_\parallel^2/s$, where $\lambda_\parallel$ is the in--plane magnetic penetration depth and $s$ the layer spacing.
However, for an infinite number of layers, the interaction between two vortices of the same sign in the same layer is logarithmic at \emph{all} length scales, not just for separations smaller than $\Lambda$.  
The difference stems from the fact that in the infinite layer problem, currents in the other layers guide a vortex's magnetic flux radially out to infinity within a disk of thickness $\lambda_\parallel$ while in the single layer problem, the flux spreads over all space.\cite{Clem91}  For two vortices of the same sign in different layers, it turns out the interaction is also logarithmic at large distances but \emph{attractive}.\cite{Artemenko90, Clem91}  Therefore, the interlayer coupling favors pancake vortices of the same sign aligning into stacks.  This is qualitatively what happens with Josephson coupling except the attractive force keeping the pancakes aligned is now logarithmic and long--ranged, instead of
linear and short--ranged.  

Because the interlayer interaction is logarithmic, it was conjectured\cite{Korshunov90} 
that the phase transition should be in the Kosterlitz--Thouless universality
class despite being a three--dimensional system.  A number of important
renormalization group studies\cite{Horovitz92, Scheidl92, Horovitz93, Timm95},
emphasizing the role of pancake vortices, explored the issue in greater detail.  In
each case, the phase transition was investigated through numerical studies of the
resulting flow equations, where the interlayer interactions were treated at varying
levels of approximation and detail (in each case, the Kosterlitz--Thouless equations
occurred as leading terms when the interlayer interaction was treated perturbatively).
Refs.~[\onlinecite{Horovitz92, Scheidl92, Horovitz93}] provided support for the KT 
scenario while Ref.~[\onlinecite{Timm95}], which was prima facie the
most complete study, reached a very different conclusion:  runaway RG flows in a very 
narrow temperature range close to $T_{KT}$ appeared to signal a three--dimensional critical
region\cite{Timm95} or perhaps a first--order transition.  This motivated the present
work.  Our central result, obtained by exactly solving the leading order RG flows which
occur when the interlayer interaction is treated non--perturbatively, is that the phase
transition is, indeed, in the Kosterlitz--Thouless universality class.   

There are a number of reasons, in addition to three--dimensionality, why
a KT transition is not a foregone conclusion for this model.  
In Ref.~[\onlinecite{Scheidl92}], it was noted that while the interlayer logarithms individually
come with much smaller coefficients than the in--plane logarithm, the infinite set of couplings
obey a ``sum rule" (see Eq.~(\ref{eq:sumrule})) which follows from flux conservation.  
Therefore, there is the possibility that the collective effect of a large number of layers 
could influence the critical properties.  In a similar vein, in Ref.~[\onlinecite{Mints00}]
it was noted that configurations involving stacks of vortices where one or both ends
terminate \emph{inside} the superconductor, which are topologically forbidden in the presence
of Josephson coupling (since vortex lines may then only terminate on the surface of the material),
should be accounted for in its absence.

In the next Section, we review some basic facts about the Biot--Savart interaction.  In Section
\ref{sec:rg} we present an analytical theory of the phase 
transition using an extension of the 2d momentum shell renormalization group\cite{Knops80}.
Accounting for configurations involving stacks of vortices, as discussed above, we re--derive
the coupled set of flow equations obtained in Ref.~[\onlinecite{Timm95}] and establish this
set as an accurate description of the low temperature physics.  We explicitly solve this set
and find that the phase transition is, in fact, in the Kosterlitz--Thouless universality class.
In Sections \ref{sec:lowT} and \ref{sec:debye}, we discuss the low and high temperature
phases, the latter using a Debye--Huckel mean field theory.  We conclude with a summary.
Technical aspects of the calculation are discussed in three appendices.  

\section{The Biot-Savart interaction}
\label{sec:bs}

In this Section, we give a physical discussion of the Biot--Savart interaction and
introduce our model Hamiltonian.  To make our assumptions clear, in Appendix \ref{app:GLtoCoulomb} we review how this interaction formally arises 
from a Ginzburg--Landau type free energy functional.  

We model our system as an infinite stack of superconducting planes, where the
stacking is in the $z$-direction and the positions of the planes are given by $z_n=ns$
where $n$ is an integer and $s$ the interlayer spacing.  A two-dimensional
``pancake" vortex\cite{Clem04} of strength $m_1$ ($m_1$ is an integer) 
placed at the origin of layer $n=0$ will induce an azimuthal
screening current $m_1K_\phi(\rho, ns)$ in layer $n$ where $\rho$ is the cylindrical radial
coordinate.  A vortex of strength $m_2$ located at position $(\rho,ns)$ will feel a radial
Lorentz force given by $F_\rho(\rho,ns)=K_\phi(\rho,ns)m_1m_2\phi_0/c$ where 
$\phi_0=\frac{hc}{2e}$.    

These screening currents were computed in Ref.~\onlinecite{Clem91} in the 
limit where $s$ is small compared to the in-plane penetration length $\lambda_\parallel$
and for distances $\rho$ large compared to $s$ and the in-plane coherence length $\xi_\parallel$:
\beq
K_{\phi}(\rho,ns=0)=\frac{c\phi_{0}s}{8\pi^{2}\lambda_{\parallel}^{2}\rho}[1-\frac{s}{2\lambda_{\parallel}}
(1-e^{-\rho/\lambda_{\parallel}})]
\label{eq:curr0}
\eeq
\beq
K_{\phi}(\rho,ns\neq 0)=-\frac{c\phi_{0}s^{2}}{16\pi^{2}\lambda_{\parallel}^{3}\rho}
(e^{-\frac{|ns|}{\lambda_{\parallel}}}-e^{-\frac{\sqrt{\rho^{2}+(ns)^{2}}}{
\lambda_{\parallel}}})
\label{eq:curr1}
\eeq
Eq.~(\ref{eq:curr0}) indicates that to leading order, the in-plane screening current $K_\phi(\rho,0)\sim \frac{1}{\rho}$.  This implies an in-plane vortex-vortex interaction potential where vortices of the same sign repel each other logarithmically with distance, as in the 2dXY model.\cite{Kosterlitz74}  If, in addition to $\rho\gg s,\xi_\parallel$, we also assume that $\rho\gg\lambda_\parallel, |ns|$, then Eq.~(\ref{eq:curr1}) indicates that the {\em out-of-plane} screening current $K_\phi(\rho,ns)\sim -\frac{1}{\rho}$.  This implies that the interaction between two vortices in different planes also varies logarithmically with distance but vortices of the same sign now {\em attract}.
Therefore, in this limit, the interaction between two vortices of strengths $m_1$ and $m_2$ at positions $(\mathbf{x_1},n_1s)$ and $(\mathbf{x_2},n_2s)$ is given by:
\beq
V_{12}\approx - q^{2} m_1 m_2 \alpha_{|n_1-n_2|}
\ln \Bigl(\frac{|\mathbf{x_1-x_2}|}{\tau}\Bigr)
\label{eq:pot}
\eeq
where
\beq
\alpha_{n}\approx\Big\{ \begin{array}{cc} -\sum_{n\neq 0} \alpha_n \approx 1-\frac{s}{2\lambda_{\parallel}}& \mbox{if $n=0$}\\
-\frac{s}{2\lambda_{\parallel}}e^{-\frac{s}{\lambda_{\parallel}}|n|} & \mbox{if $n\neq 
0$}\end{array}
\label{eq:interactions}
\eeq
, $q=\sqrt{\frac{\phi_{0}^{2}}{8\pi^{2}s}(\frac{s}{\lambda_{\parallel}})^{2}}$, and
$\tau$ is a nominal short distance length scale (of order $\lambda_\parallel$).
 
Eq.~(\ref{eq:pot}) is a model of the Biot-Savart interaction which was considered in 
Refs.~\onlinecite{Scheidl92} and \onlinecite{Timm95}, and is the model considered in
this paper.  A noteworthy feature of this interaction is that 
the coupling constants obey a ``sum rule":
\beq
\sum_{n} \alpha_{n}=0
\label{eq:sumrule}
\eeq
This is not an accident but follows from flux quantization\cite{Scheidl92} and
is related to an important feature of the full interaction (Eqs.~(\ref{eq:curr0}), (\ref{eq:curr1}), and (\ref{eq:Vmn})):  the current distribution of an infinite stack of pancake vortices, one in each layer, is 
exponentially screened at distances $\rho \gg \lambda_\parallel$,\cite{Clem91} in analogy with the well-known result for a vortex line in a bulk three--dimensional superconductor.\cite{Tinkham96}  In contrast, for a stack of uncoupled two--dimensional layers, the current in each layer would decay according to the Pearl
criterion\cite{Pearl64}:  as $\sim \frac{1}{\rho}$ for distances $\rho < \Lambda$ and as $\sim \frac{1}{\rho^2}$ for $\rho\gg\Lambda$, where $\Lambda = \frac{2\lambda_\parallel^2}{d}$, $d$ being
the layer thickness.

The electromagnetic interaction causes (same charge) pancake vortices in different layers to preferentially align into stacks, which is phenomenologically what happens with Josephson coupling except that now the aligning force is logarithmic and long-ranged, as opposed to linear and between neighboring layers in the Josephson coupled case.  In particular, even though $\alpha_1\ll\alpha_0$, which naively suggests that only the in-plane interaction is important, the $\alpha$'s decay very slowly and the sum rule indicates that the combined effect of {\em many} layers could possibly lead to three-dimensional effects.  We will return to this issue in the next Section.  

We conclude this Section by discussing some of the approximations inherent in Eq.~(\ref{eq:pot}).
In the high temperature superconductors, typical orders of magnitude\cite{Clem91} are $s\approx 12$ $\buildrel _{\circ} \over {\mathrm{A}}$ and $\lambda_\parallel\approx 1400$ $\buildrel _{\circ} \over {\mathrm{A}}$ so the smallness of $\frac{s}{\lambda_\parallel}\approx 10^{-2}$ assumed in the derivation is met in practice.  In and near the low temperature superconducting phase, we expect the density of (thermally--excited) pancake vortices in each layer to be small and hence the characteristic vortex-vortex separation $\bar{\rho}$ to be large.  In our analysis, we assume that $\bar{\rho}$ is large compared to other characteristic lengths, such as $\lambda_\parallel$, which is one of the conditions for the interlayer logarithmic form to be valid.  Also, in an infinite layer system, the logarithmic approximation will break down for interlayer separations $|ns|\approx\bar{\rho}$.  However, the length scale associated with the convergence of the sum rule (Eq.~(\ref{eq:sumrule})) is of order $\lambda_\parallel$ so, if $\bar{\rho}\gg\lambda_\parallel$, we expect 
the net contribution of these farthest layers to be small regardless of whether the full (Eq.~(\ref{eq:Vmn})) or simplified (Eq.~(\ref{eq:pot})) interaction is used.  Therefore, in this paper, we will approximate the Biot-Savart interaction by its long distance form, which permits us to use Coulomb gas techniques to analyze the partition function.  However, we will retain some aspects of the short distance physics, i.\ e.\ on length scales shorter than $\lambda_\parallel$, by introducing fugacity variables as discussed in the next Section.

\section{Renormalization group analysis}
\label{sec:rg}

The partition function of a layered superconductor, where each layer contains a neutral gas
of thermally--excited pancake vortices is:
\begin{equation}
\mathcal{Z}=\sum_{\{N_{k,l}\}}\prod_{k,l}y_{k,l}^{N_{k,l}}\sum_{\{c\}} \exp(-\beta\sum_{i\neq j} \tilde{V}_{ij}),
\label{eq:Z1}
\end{equation}
where $y_{k,l}=\exp(-\beta E_{k,l})$, $\beta$ being the inverse temperature and 
$E_{k,l}$ the energy cost of creating a pancake vortex of type $k$ in layer $l$, the species label $k$ denoting both strength and sign.  We assume the layers are equivalent 
which implies the fugacities are the same in each layer, i.e.\ $y_{k,l}=y_k$.  $N_{k,l}$ is the number
of type $k$ vortices in layer $l$.  The sum on $\{N_{k,l}\}$ is over layer occupations which satisfy
charge (vortex) neutrality in each layer.  The sum on $\{c\}$ is over spatial configurations of vortices consistent with the set $\{N_{k,l}\}$.\footnote{A common way of writing Eq.~(\ref{eq:Z1}) is with the sum over $\{c\}$ replaced by a multidimensional integral over the coordinates of the vortices\cite{Kosterlitz74,Timm95} in which case factors $1/(N_{k,l})!$ are needed because the vortices are indistinguishable.}  $\tilde{V}_{ij}$ is the vortex--vortex interaction
which includes a hard--core constraint that two vortices in the \emph{same} layer must be 
separated by a distance $\tau$ of order $\xi_{\parallel}$, the in--plane coherence length. 

At this stage, $\tilde{V}_{ij}$ is the exact vortex--vortex interaction.  To make analytical
progress, we separate $\sum_{i\neq j} \tilde{V}_{ij}$ into two parts, $\sum^{<}_{i\neq j} \tilde{V}_{ij}
+\sum^{>}_{i\neq j} \tilde{V}_{ij}$, corresponding to the contribution from pairs of vortices with
in--plane separation $\rho<\lambda_{\parallel}$ and $\rho>\lambda_{\parallel}$ respectively.
The latter interaction is given by Eq.~(\ref{eq:pot}) and in a dilute system will apply for most of the pairs
(with the caveats mentioned the previous Section).  We approximate the shorter--distance physics
in two steps.  The possibility of having two vortices in the same plane separated by a
distance less than $\lambda_\parallel$ can be accounted for by suitably redefining the fugacity variables.\footnote{More precisely, the core energy $E_{k,l}$ will no longer represent a ``bare" core energy but a renormalized energy that includes the effects of screening due to closely spaced vortex--antivortex pairs.  This screening will also renormalize the coefficients $\{\alpha_n\}$ of the logarithmic long--distance interaction from the values in Eq.~(\ref{eq:interactions}).  For a sufficiently dilute system of vortices, we expect these renormalizations to be very small.  Therefore, we will still use the values in Eq.~(\ref{eq:interactions}) as initial conditions for the RG analysis of Eq.~(\ref{eq:Z2}).}  To 
approximate the interaction between two closely spaced vortices in different layers, we introduce a new set of fugacity variables $\{ w_{ab; ij} \}$, where $w_{ab;ij} = \exp(-\beta E_{ab;ij})$, $E_{ab;ij}$ being the interaction energy of having a pancake vortex of type $b$ in layer $j$ directly above a pancake vortex of type $a$ in layer $i$.  For equivalent layers, $w_{ab;ij}$ will depend only on $|i-j|$ (and the strengths $a$ and $b$).  With these approximations,
Eq.~(\ref{eq:Z1}) becomes:
\begin{multline}
\mathcal{Z}=\\ \sum_{\{N_{k,l};N_{ab;ij}\}}\prod_{k,l}y_{k,l}^{N_{k,l}}\prod_{ab;ij}w_{ab;ij}^{N_{ab;ij}}\sum_{\{ c \} } \exp(-\beta{\sum_{i\neq j}}^{>} V_{ij}),
\label{eq:Z2}
\end{multline}
where $N_{ab;ij}$ is the number of pairs of vortices where the first member is a vortex of type $a$ in layer $i$ which is directly below the second member, a vortex of type $b$ in layer $j$.  The sum over $\{ c\}$ is now over those spatial configurations of pancakes consistent with a set of (charge--neutral) layer occupations $\{N_{k,l}\}$ and interlayer patterns $\{ N_{ab;ij} \}$.  The sum inside the exponential is over pairs of vortices that are laterally separated by at least $\lambda_\parallel$ and their interaction
$V_{ij}$ is given by Eqs.~(\ref{eq:pot})--(\ref{eq:interactions}).  

There is another way of viewing Eq.~(\ref{eq:Z2}), which is more in-line with previous
treatments of a vector Coulomb gas.  Two vortices in the same layer are always separated by a distance of at least $\lambda_\parallel$, which may then be viewed as the effective ``size" of a vortex.  For a system with an infinite number of layers, each having a small but nonzero \emph{density} of vortices, an infinitely long cylinder of radius $\lambda_\parallel$ perpendicular to and piercing the layers will ``catch" an infinite number of vortices.  The stack of pancakes caught by
the cylinder may be viewed as an extended object.  By densely packing the system with such cylinders, the configurations of the system may be viewed as configurations of these extended objects.  
The extended objects can be labelled by a species index $\mathbf{n}=(\dots,n_1,n_2,\dots)$ where $n_i$ is an integer indicating the strength and sign of the vortex occupying layer i of the object in question (for a dilute system, most of the entries in $\mathbf{n}$ will be zero).  We can formulate the problem
in terms of these extended objects in which case the partition function Eq.~(\ref{eq:Z1})
becomes:
\begin{equation}
\mathcal{Z}=\sum_{\{\mathbf{n}\}}\prod_{\mathbf{n}}y_{\mathbf{n}}^{N_{\mathbf{n}}}\sum_{\{c\}} \exp(-\beta\sum_{i\neq j} {V}_{\mathbf{n}_i,\mathbf{n}_j}),
\label{eq:Z3}
\end{equation}
where $\sum_{\{\mathbf{n}\}}$ is a sum over sets of extended objects consistent with charge 
neutrality in each layer and $\sum_{\{c\}}$ is over distinct spatial configurations of these objects.
The interaction between two of these objects, indexed by $\mathbf{n}_i$ and $\mathbf{n}_j$, is logarithmic by construction and given by the sum of the pairwise interactions between the pancakes comprising each stack:
\begin{equation}
V_{\mathbf{n}_i,\mathbf{n}_j}= \sum_{k,l} V_{n_{ik},n_{jl}}
\label{eq:pairs}
\end{equation}
where $k,l$ are layer indices and $V_{n_{ik},n_{jl}}$ is given by Eq.~(\ref{eq:pot}).   Since 
an extended object is composed of an infinite number of pancakes, it will require an infinite creation energy $E_\mathbf{n}$ and the corresponding fugacity $y_\mathbf{n}=\exp(-\beta E_\mathbf{n})$ will be formally zero.\footnote{There is no inconsistency because any 
physical quantity will involve an average where a substantial part of Eq.~(\ref{eq:Z3}) will
appear in both the numerator and denominator and hence cancel.}    However, we can
write the energy of an extended object as:
\begin{equation}
E_\mathbf{n}=\sum_k E_{n_k,k} + \sum_{k\neq l} E_{n_kn_l; kl}
\label{eq:core}
\end{equation}
where the first term is the creation energy of each pancake in the stack while the second
is the pairwise  interaction energy between different pancakes in the same stack.
Eqs.~(\ref{eq:pairs}) and (\ref{eq:core}) show that Eqs.~(\ref{eq:Z2}) and (\ref{eq:Z3}) are
equivalent ways of expressing the partition function.

\begin{figure}[ht]
{\begin{center}
\includegraphics[width=2in]{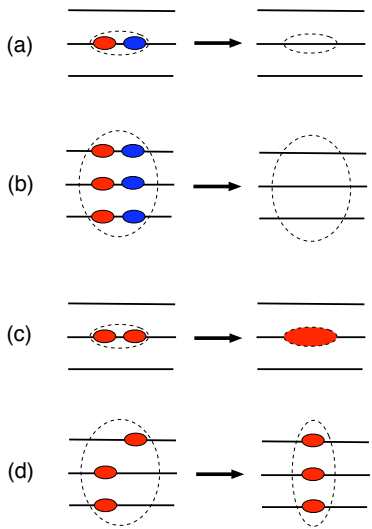}
\caption{Pictorial summary of the renormalization group procedure 
discussed in Appendices \ref{app:duality} and \ref{app:flow}.  If 
an object and its anti-object are very closely spaced, then they will 
cancel upon coarse graining, but in the process the coupling 
constants will be renormalized.  (a) shows the simplest such process 
involving single pancakes which may be viewed as integrating out the 
smallest loops of magnetic flux.  (b) is an example of the
more complicated object/anti--object cancellations which correspond 
to integrating out larger loops.  Nearby objects that are not antiparticles
will fuse together.  (c) shows two single strength pancakes fusing into a double strength 
pancake.  (d) shows a height two stack and a single pancake fusing into a 
height three stack.}
\label{fig:coarse}
\end{center}}
\end{figure}

Our goal is to determine the phase diagram of the model described by Eq.~(\ref{eq:Z2})
(or Eq.~(\ref{eq:Z3})) in the dilute limit where each plane contains a small, but nonzero,
density of vortices.  The advantage of separating the vortex--vortex interaction into
a long--distance logarithmic term, accounting for the short--distance physics through
generalized fugacity variables, is that it permits the use of renormalization group (RG) 
techniques developed for studying the two--dimensional Coulomb gas.\cite{Kosterlitz74,
Knops80, Nienhuis87}  The procedure requires us to consider a generalized version
of Eq.~(\ref{eq:Z2}):
\begin{multline}
\mathcal{Z}=\\ \sum_{\{N_{k,l};N_{ab;ij}\dots\}}\Bigl(\prod_{k,l}y_{k,l}^{N_{k,l}}\prod_{ab;ij}w_{ab;ij}^{N_{ab;ij}}\prod_{abc;ijk} w_{abc;ijk}^{N_{abc;ijk}}
\\\prod_{abcd;ijkl} w_{abcd;ijkl}^{N_{abcd;ijkl}}\dots\Bigr)
 \sum_{\{ c \} } \exp(-\beta{\sum_{i\neq j}}^{>} V_{ij}),
\label{eq:Z4}
\end{multline}
In this equation, $w_{abc;ijk} = \exp(-\beta E_{abc;ijk})$ where $E_{abc;ijk}$ is the 
three--body interaction energy of an aligned triplet of vortices where the first member
is a vortex of type $a$ in layer $i$, directly below the second member, a vortex of
type $b$ in layer $j$, which is directly below the third member, a vortex of type $c$
in layer $k$.  For equivalent layers, $E_{abc;ijk}$ will depend only on $|i-j|$ and 
$|j-k|$ (and the strengths $a$, $b$, and $c$).  Similarly, we include fugacity variables
corresponding to four (and higher) body interactions.  These terms may be
visualized in the extended object picture where Eq.~(\ref{eq:core}) generalizes
to:
\begin{equation}
E_\mathbf{n}=\sum_k E_{n_k,k} + \sum_{k\neq l} E_{n_kn_l; kl} +\sum_{k<l<m} E_{n_kn_ln_m;klm} + \dots
\label{eq:core2}
\end{equation}
In Appendices \ref{app:duality} and \ref{app:flow}, we present a detailed renormalization
group treatment of this model.  The analysis may be viewed as an iterative coarse--graining
procedure connecting our model with a series of other models with the same critical properties.  
A physical picture of this procedure is illustrated and discussed in Fig.~\ref{fig:coarse}.

The analysis of Appendix \ref{app:flow} yields an infinite set of coupled flow equations
for the extended object fugacities $\{ y_{\mathbf{n}} \}$ and the analogous equations
for the $(\{ y_{k,l} \}, \{w_{ij;kl} \}, \dots)$ variables.  The system has a fixed point when 
the $\{ y_{\mathbf{n}} \}$ variables are identically zero.  A linearized theory about this 
fixed point (Eq.~(\ref{eq:flowyn})) suggests that at sufficiently low temperatures the 
fixed point is stable while at higher temperatures these variables become RG relevant.
Beginning in the low temperature phase and raising the temperature, the phase transition
is indicated by one of the $\{ y_{\mathbf{n}} \}$ becoming marginal.  Which fugacity is the
first to ``unbind" depends on the initial conditions of the RG flow.  

If the starting model is given by Eq.~(\ref{eq:interactions}), then the first extended objects 
to become marginal are those where $\mathbf{n}$ has +1 in one of its entries and zeroes 
everywhere else.  By Eq.~(\ref{eq:core2}), the fugacity of this object is precisely the fugacity
of a single strength pancake $y$ ($\equiv y_{1,i}$ but since we have assumed a translationally
invariant system, the layer index is not needed).  The flow equation for this quantity, in the 
approximation where we keep only the leading $y$ dependence, is:
\beq
\frac{dy^{2}}{d\epsilon}=y^{2}(4-\beta q^{2}\alpha_{0}).
\label{eq:y}
\eeq
In the same limit, the flow equations for the coupling constants are:
\beq
\frac{d(\beta q^{2}\alpha_{n})}{d\epsilon}=-\pi y^{2}\sum_{m}(\beta q^{2}
\alpha_{n-m})(\beta q^{2}\alpha_{m}).
\label{eq:a}
\eeq
where we have also taken the ``distance to marginality'', $(4-\beta q^2\alpha_0)$, as a 
small parameter.  We will discuss the flow equations for the other variables further
below.

Eqs.~(\ref{eq:y}) and (\ref{eq:a}) are precisely the flow equations obtained in 
Ref.~\onlinecite{Timm95}, using a formulation that involved only single pancakes,
without accounting for extended objects.  Our analysis implies that the more 
complicated objects are irrelevant at temperatures below the transition so in this limit,
Eqs.~(\ref{eq:y}) and (\ref{eq:a}) may have greater validity than initially 
suspected.  As discussed in the Appendix \ref{app:flow}, this irrelevance does not simply follow
from the linearized theory but involves using the sum rule (Eq.~(\ref{eq:sumrule})) to
place a bound on higher order terms -- a technical issue that does not arise in the
single--layer problem.

We continue our analysis by recognizing that the right side of 
Eq.~(\ref{eq:a}) is a convolution of the couplings.  Taking the Fourier 
transform, we obtain:
\beq
\frac{d(1/(\beta q^{2}\alpha(k)))}{d\epsilon}=\pi y^{2}
\label{eq:a2}
\eeq
where $\alpha(k)=\sum_{n} \alpha_{n} e^{-ikn}$ and we used the fact 
that $\alpha_{m}=\alpha_{-m}$.  Because the right 
side is independent of $k$, we may formally integrate this equation to 
obtain:
\beq
\beta q^{2}\alpha(k,\epsilon)=\frac{\beta q^{2}\alpha(k,0)}
{1+\beta q^{2}\alpha(k,0)C(\epsilon)}
\label{eq:a3}
\eeq
where $C(\epsilon) \equiv \int_0^\epsilon \pi y^2$ is an integration 
constant that obeys the flow equation:
\beq
\frac{dC}{d\epsilon}=\pi y^{2}
\label{eq:C}
\eeq
with the initial condition $C(0)=0$. Observe that $\alpha(0,\epsilon)=\sum_{n}\alpha_{n}$ 
so Eq.~(\ref{eq:a3}) implies that the sum rule is preserved by the flow.

\begin{figure}[ht]
{\begin{center}
\includegraphics[width=3in]{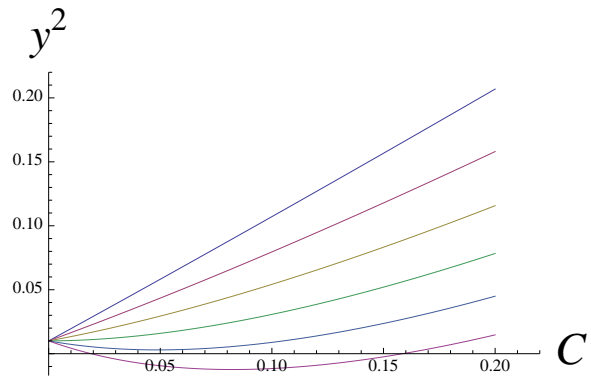}
\caption{Plot of the flow trajectories of Eq.~(\ref{eq:traj}) for the 
case where the initial fugacity $y_{0}=0.1$.  The six curves, from
top to bottom, are for $\beta q^2=1$, 2, 3, 4, 5, and 6 respectively.
Since the flow stops when $y=0$, this initial condition implies a 
phase transition occurring for $\beta q^2$ between 5 and 6.}
\label{fig:traj}
\end{center}}
\end{figure}

Using Eqs.~(\ref{eq:y}) and (\ref{eq:C}), we obtain a differential 
equation for the flow trajectories in the $(C,y)$-plane, which can
be integrated:
\bea
y^{2}=y_{0}^{2}+\frac{1}{\pi}\Bigl(4C-
\int_{-\pi}^{\pi}\frac{dk}{2\pi}
\ln(1+\beta q^{2}|\alpha(k,0)C(l)|)\Bigr)\nonumber\\
\label{eq:traj}
\eea
Representative trajectories are sketched in Fig.~\ref{fig:traj} for a given 
value of $y_{0}$ as a function of temperature.  As $C$ increases from 0 
monotonically during the flow (Eq.~(\ref{eq:C})), the system moves along
these curves from left to right.  At high temperatures, the fugacity increases
monotonically during the flow.  As the temperature is lowered, the fugacity
initially decreases before increasing.  At a critical temperature, the curve 
will intersect the $y=0$ axis at one point and at lower temperatures, the
curves cross the axis.  However, $y=0$ is a fixed point of Eqs.~(\ref{eq:y})
and (\ref{eq:a}), which means the system will ``stop" once $y=0$ is 
reached.  These curves demonstrate the existence of a low temperature phase, 
where the fugacity renormalizes to zero, separated from a high temperature region 
by a transition corresponding to the unbinding of 2$d$ pancake vortices: 
this, by definition, is a Kosterlitz-Thouless transition.\cite{Kosterlitz74}

The most distinguishing characteristic of the single--layer KT transition is the universal
jump in the superfluid stiffness.\cite{Nelson77}  To see what happens
in the layered case, we need to determine the range of initial conditions for which
the trajectory defined by Eq.~(\ref{eq:traj}) passes through $y=0$ at some 
$C>0$.  That is, we need to solve:
\beq
\pi y_0^2 = \int \frac{dk}{2\pi} \ln (1+\beta q^2 |\alpha(k,0)| C) - 4C 
\eeq
If we begin close to the critical point, and assume the position of this point will only be 
changed by a small amount relative to the single--layer case, we can expand this expression 
treating $y_0$, $C$, and $(\beta q^2\alpha_0 - 4)$ as small parameters.  Then, to leading
order, we obtain:  
\beq
\pi y_0^2 \approx(\beta q^2\alpha_0 (0)-4) C - \frac{1}{4} \sum_m \alpha^2_m (0) C^2 
\eeq
where we have inverted the Fourier transform.  In order to have a $y=0$ solution where 
$C>0$, the following criterion must be satisfied:
\beq
\beta q^2\alpha_0(0)-4 > \sqrt{ \pi y_0^2\sum_m \alpha^2_m(0)}
\eeq
Therefore, at temperatures $T>q^2\alpha_0(0)/4$, there is no solution and $y$ will diverge
as the flow coordinate $\epsilon\rightarrow\infty$.  Eq.~(\ref{eq:a3}) indicates that the couplings 
$\{ \alpha_n \}$ will go to zero in this same limit.  The critical temperature at which a $y=0$
fixed point exists is $T=q^2\alpha_0(0)/4$.  This solution represents a critical surface $(y=0,
\{\alpha_n=\alpha_n(0)\})$ where the only constraints on the couplings $\{ \alpha_n(0) \}$ are
the sum rule (or, more precisely, that the matrix $\alpha_{ij}\equiv\alpha_{|i-j|}$ is positive definite --- 
see the discussion in Appendix \ref{app:duality}) and that the values are such that the single
strength pancake is the first fluctuation to become marginal.    

To relate this to the universal jump, we need to relate the quantity $\alpha_0(0)$, given in Eq.~(\ref{eq:interactions}), to the in--plane superfluid stiffness measured in an experiment.  
The in--plane superfluid stiffness is defined in terms of the in--plane magnetic penetration depth $\lambda_{\parallel}$, which is a directly measurable quantity: $\rho_s=\phi_0^2s/(16\pi^3\lambda_\parallel^2)=q^2/2\pi$ (see the discussion between Eqs.~(\ref{eq:LD}) and (\ref{eq:F})).  As the critical temperature is crossed, the quantity $\beta q^2 \alpha_0(\infty)$ jumps downward from 4 to zero.  In terms of $\rho_s$:
\beq
\left [ \frac{\rho_s\alpha_0(\infty)}{T} \right ]_{T_c^-}^{T_c^+} = \frac{2}{\pi} \label{eq:jump1}
\eeq
\beq
\frac{\left [\rho_s\right ]_{T_c^-}^{T_c^+}}{T_c} = \frac{2}{\pi \left (1-\frac{s}{2\lambda_\parallel}\right)}
\label{eq:jump2}
\eeq
Therefore, while the jump described by Eq.~(\ref{eq:jump1}) is a universal quantity, the jump in the
superfluid stiffness (Eq.~(\ref{eq:jump2})), which is the quantity that is directly measured, receives a 
non--universal correction on the order of 1 percent for Bi$_2$Sr$_2$CaCu$_2$O$_{8+x}$.

\section{Low temperature phase}
\label{sec:lowT}

Having established the existence of a critical point and phase transition, we now turn to the
nature of the low temperature phase and the way the present case differs from a stack of decoupled
layers.  The most direct difference follows from Eq.~(\ref{eq:a3}).  As the low temperature phase is 
characterized by a finite value of $C(\infty)$, Eq.~(\ref{eq:a3}) indicates that the interlayer couplings
$\{ \alpha_n \}$ will have nonzero fixed point values.    Therefore, interlayer correlations will be 
present.

Another perspective may be gained by considering the flow equations for the pair interaction
fugacity for single strength pancakes in layers 1 and $m$, which we label $w_{1m}$ (Eq.~(\ref{eq:RGw1m})):
\begin{equation}
\frac{dw_{1m}}{d\epsilon} = -\left [ 2+\beta q^2 \alpha_{m-1} \right ] w_{1m} - \beta q^2 \alpha_{m-1}
\label{eq:w1m}
\end{equation}
where we have dropped terms that flow to zero in the low temperature phase as $\epsilon\rightarrow\infty$.
Eq.~(\ref{eq:w1m}) has a fixed point when:
\begin{equation}
w_{1m}(\infty) = \frac{\beta q^2 \left |\alpha_{m-1}(\infty)\right |}{2+\beta q^2 \left |\alpha_{m-1}(\infty)\right |}\approx 
\frac{\beta q^2 \left |\alpha_{m-1}(\infty)\right |}{2}
\label{eq:w1m-1}
\end{equation}
In the decoupled limit, $w_{1m}(\infty)$ would be zero.  The $y=0$ fixed point means that at
the longest length scales, the effective model is equivalent to one where the vortices are not
present:  i.\ e.\ the properties of the superconductor will be governed by the spin wave part of the
action (Eq.~(\ref{eq:swvort})).  However, at intermediate length scales, which correspond to 
the points forming the RG trajectory, the effective model will be one where the vortices are present, albeit with small
fugacity.  $w_{1m}$ flowing to zero means that configurations where vortices lie on top of one
another contribute progressively less to the partition function, relative to configurations where they do not, as the system is viewed from larger length scales.  The reason is purely entropic.\footnote{A coarse--graining step may be viewed as taking the effective core size of a vortex from $\tau$ to $\tau+d\tau$.  A vortex in the new model represents $(1+\frac{d\tau}{\tau})^2$ configurations of the old model.
This is accounted for by renormalizing the fugacity.  Now consider an extended object $y_{1m}$, with pancakes in layers 1 and $m$.  In the coarse--grained model, this object will still represent $(1+\frac{d\tau}{\tau})^2$ configurations of the old model but merely increasing the size of the pancakes will give an extra factor of $(1+\frac{d\tau}{\tau})^2$.  Hence, because $y_{1m}=y^2w_{1m}$, we can account for this overcounting by renormalizing $w_{1m}$ by the factor $(1-\frac{d\tau}{\tau})^2$.}  Eq.~(\ref{eq:w1m-1}) indicates an enhanced probability for such configurations in the coupled layer case.  

At one level, this is not so surprising because at the smallest length scales, the interlayer
couplings (Eq.~(\ref{eq:interactions})) favor the formation of stacks.  However, it is interesting
to discuss the higher--body terms.  For example, we may obtain the three-body interaction fugacity for single strength pancakes in layers $a$, $b$, and $c$, which we label $w_{abc}$.  To obtain this quantity, it is easiest to start with the flow equation for $y_\mathbf{n}\equiv y_{abc}$ where $\mathbf{n}$ is a vector with $+1$ in layers $a$, $b$, and $c$, and 0 elsewhere.  From Appendix \ref{app:flow}, it follows, to leading order in $y$,
that:
\begin{multline}
\frac{dy_{abc}}{d\epsilon}=\left [ 2-\frac{\beta q^2}{2}\left (3\alpha_0+\alpha_{ab}+\alpha_{ac}+\alpha_{bc}\right)\right] y_{abc}\\ - y\Bigl [y_{ab}(\alpha_{ac}+\alpha_{bc}) + y_{ac}(\alpha_{ab}+\alpha_{bc})+y_{bc}(\alpha_{ab}+\alpha_{ac})\Bigr ]
\label{eq:yabc}
\end{multline}
where $y_{ij}$ is the fugacity of the extended object with single strength pancakes in layers $i$ and $j$, the other layers being empty, and $\alpha_{ij}\equiv\alpha_{|i-j|}$.  We can determine the low temperature fixed point value of $y_{abc}$ by setting the right side of Eq.~(\ref{eq:yabc}) to zero.  By Eq.~(\ref{eq:core2}), these fugacities may be understood as a product of terms associated with the creation and interaction of the stacked vortices:  $y_{ab}=y^2 w_{ab}$ and $y_{abc}= y^3 w_{ab}w_{ac}w_{bc}w_{abc}\equiv y^3 v_{abc}$.  Using Eq.~(\ref{eq:w1m-1}), we find:
\begin{multline}
v_{abc}(\infty)\\ =\frac{\frac{\beta q^2}{2}\left [ (\alpha_{ab}+\alpha_{ac}+\alpha_{bc})^2-(\alpha_{ab}^2+\alpha_{ac}^2+\alpha_{bc}^2)\right]}{(\frac{3\beta q^2}{2}\alpha_0-2)+\frac{\beta q^2}{2}\left (\alpha_{ab}+\alpha_{ac}+\alpha_{bc}\right)}\Biggl |_{\infty} \\ \approx \frac{\frac{\beta q^2}{2}\left [ (\alpha_{ab}+\alpha_{ac}+\alpha_{bc})^2-(\alpha_{ab}^2+\alpha_{ac}^2+\alpha_{bc}^2)\right]}{\left(\frac{3\beta q^2}{2}\alpha_0-2\right)}\Biggl |_{\infty}
\label{eq:vabc}
\end{multline}
In terms of the variable $w_{abc}$, the result is:
\begin{multline}
w_{abc}(\infty)\\ \approx \frac{\left(\frac{2}{\beta q^2}\right)^2\left [ (\alpha_{ab}+\alpha_{ac}+\alpha_{bc})^2-(\alpha_{ab}^2+\alpha_{ac}^2+\alpha_{bc}^2)\right]}{\alpha_{ab}\alpha_{ac}\alpha_{bc} \left(\frac{3\beta q^2}{2}\alpha_0-2\right)}\Biggl |_{\infty}
\label{eq:wabc}
\end{multline}
In decoupled limit, $v_{abc}$ and $w_{abc}$ flow to zero and infinity respectively, which again is
purely entropic.  $v_{abc}$ being nonzero indicates an enhanced probability for these objects
in the coupled layer case.  However, these expressions indicate that the effective interaction
between the vortices within the stack is no longer a simple pairwise form.  Qualitatively, this is
another indication of three--dimensional correlations being present in the low temperature
phase.  

\section{High temperatures}
\label{sec:debye}

Fig. \ref{fig:traj} shows that the high temperature phase has $C\rightarrow\infty$ as $\epsilon\rightarrow\infty$.  From Eq.~(\ref{eq:a3}), we conclude that the new fixed point model is one where all of the couplings have renormalized to zero.  In the same limit, $y$ also diverges, which may be interpreted as an unbinding of pancake vortices.  Therefore, the high temperature phase may be viewed as a stack of nearly independent planes, each containing a neutral plasma of weakly interacting vortices.  However, once $y$ becomes of order unity, the renormalization group approach discussed in the previous Section no longer applies.  The flow equations (\ref{eq:y}) and (\ref{eq:a}) were obtained by dropping terms involving higher powers of $y$, which is no longer valid when $y$ is order unity.  Physically, the RG approach of Appendix \ref{app:flow} assumes a dilute gas of vortices, which is no longer true at high temperatures.

Therefore, to gain insight into the nature of the high temperature phase, a different analytical
approach is needed.  We study this limit using a Debye--Huckel mean field analysis where we 
assume each layer has $N$ positive and $N$ negative vortices, which we represent via density 
functions $\rho^{+}_m(\mathbf{x})$ and $\rho^{-}_m(\mathbf{x})$, where $m$ is a layer
index and $\mathbf{x}$ is the in--plane coordinate.  The total charge density in
layer $m$ is given by $\rho_m(\mathbf{x)}=\rho^+_m(\mathbf{x})-\rho^-_m(\mathbf{x})$.  The 
system is modeled by the mean--field free energy functional:
\begin{multline}
\mathcal{F}=\frac{1}{2}\sum_{mn}\left[\int\int d^{2}x d^{2}y \rho_m(\mathbf{x})\rho_n(\mathbf{y})
 V_{mn}(\mathbf{x}-\mathbf{y})\right ]
+\\ T\sum_{m}\left[\int 
d^{2}x \left(\rho^{+}_m(\mathbf{x})\ln(\frac{\rho^{+}_m(\mathbf{x})}{N}))+\rho^{-}_m(\mathbf{x})\ln(\frac{\rho^{-}_m(\mathbf{x})}{N})\right)\right]
\\ +\sum_m \int d^2x \phi^{\text{ext}}_m(\mathbf{x})\rho_m(\mathbf{x})\\
\label{eq:varfree}
\end{multline}
where the three terms are the vortex--vortex interaction, system entropy, and interaction of
the vortices with an external potential $\phi^\text{ext}_m(\mathbf{x})$.  The potential 
$V_{mn}(\mathbf{x}-\mathbf{y})$ is given by Eq.~(\ref{eq:pot}).  The idea is to minimize
$\mathbf{F}$ with respect to the functions $\rho^{\pm}_m(\mathbf{x})$ subject to the 
constraint: $\int d^2x \rho^{\pm}_m(\mathbf{x}) = N$. 

The previous paragraph is a natural way to discuss 
the high temperature limit of a layered Coulomb gas interacting via Eq.~(\ref{eq:pot}).\cite{Chaikin95}  However, there is a caveat to note when we relate this model to superconductors.  The phase transition discussed in the previous Section describes a loss of superconductivity, which is the low temperature state, due to a loss of long--range order in the \emph{phase} of the order parameter.  Therefore, in principle, one may have a regime without superconductivity but where the \emph{amplitude} of the superconducting 
order parameter is nonzero.\cite{Emery02}  It is in such a regime that our model applies since it is reasonable to assume the basic degrees of freedom would still be vortices.  A nonzero order parameter amplitude means the effective penetration depth is finite so we expect the interlayer mechanism of Eq.~(\ref{eq:pot}), which
is ultimately due to the screening currents, to still apply.  On the other hand, once the order parameter
amplitude is zero, we can no longer think of the basic degrees of freedom as vortices so our model would no longer be relevant.  In Ref.~[\onlinecite{Emery02}], it was suggested that for layered superconductors with small superconducting carrier densities, including the high--$T_c$ materials, the energy scale associated with phase decoherence might be appreciably less than the energy scale at which the amplitude goes to
zero.  In such a case, there would be a temperature range above $T_c$ where our model would apply.

In the absence of an external potential, the minimum free energy is obtained when both the 
positive and negative charges are uniformly distributed in each layer, i.e. $\rho^{\pm}_m(\mathbf{x})=N/A\equiv\rho_0$ where $A$ is the area of a layer.  If we perturb the system about this limit with a small $\phi^\text{ext}$, the corresponding density fluctuation may be calculated as a linear response:
\begin{equation}
\delta\rho_m(\mathbf{x}) = -\sum_n \int d^2x' \chi_{mn}(\mathbf{x}-\mathbf{x'}) \phi^\text{ext}_n(\mathbf{x'})
\end{equation}
where the (Fourier transform) of the susceptibility $\chi_{mn}(\mathbf{x},\mathbf{x'})\equiv -(\delta\rho_m(\mathbf{x})/\delta\phi^\text{ext}_n(\mathbf{x'}))_{\phi=0}$ is given by:
\begin{equation}
\chi(\mathbf{q},k)=\frac{1}{\frac{T}{2\rho_{0}}+V(\mathbf{q},k)}
\label{eq:chik}
\end{equation}
where $V(\mathbf{q},k)=\alpha(k)\frac{2\pi}{q^{2}}$ 
is the Fourier transform of Eq.~(\ref{eq:pot}) and 
$\alpha(k)$ the Fourier transform of Eq.~(\ref{eq:interactions}):
\bea
\alpha(k)&=&\sum_{n}\alpha(n)e^{-ikn}=\alpha_{0}+\sum_{n\neq 
0}\alpha_{n}e^{-ikn}\nonumber\\&=&\frac{s}{2\lambda_\parallel}
\Bigl[\frac{e^{s/\lambda_\parallel}+1}{e^{s/\lambda_\parallel}-1}\Bigr]
\Bigl[\frac{\cos k-1}{\cos k-\cosh \frac{s}{\lambda_\parallel}}\Bigr]
\eea
We choose $\phi^\text{ext}(\mathbf{x})$ to be the potential of a 
unit test charge at the origin of the $n=0$ layer:  
$\phi^{\text{ext}}(\mathbf{q},k)=\frac{2\pi}{q^{2}}\alpha(k)$   While this potential
is not small near the origin, our interest is in the long distance 
behavior.  The 
corresponding density fluctuation is:
\begin{multline}
\delta\rho(\mathbf{q},k)=
-\chi(\mathbf{q},k)\phi^{\text{ext}}(\mathbf{q},k)=
\frac{\eta^{2}}{q^{2}+\eta^{2}}\left[\frac{\cos k-1}{\cos 
k-a_{q}}\right ] \\
\end{multline}
where $\eta^{2}=\frac{4\pi\rho_{0}}{T}
\frac{s}{2\lambda_\parallel}\Bigl[\frac{e^{s/\lambda_\parallel}+1}{e^{s/\lambda_\parallel}-1}\Bigr]$ 
and $a_{q}=\frac{q^{2}\cosh\frac{s}{\lambda_\parallel}+\eta^{2}}{q^{2}+\eta^{2}}$.
Taking the inverse Fourier transform in the layering direction:
\begin{multline}
\delta\rho_m(\mathbf{q})=\frac{\eta^{2}}{q^{2}+\eta^{2}}
\left[\delta_{m,0}-\sqrt{\frac{a_{q}-1}{a_{q}+1}}\left(a_q-\sqrt{a_q^2-1}\right )^m\right ] \\
\label{eq:drhoq}
\end{multline}
The inverse transform in the $\mathbf{q}$ direction is not a simple 
expression.  However, if we expand in the small parameter 
$\frac{s}{\lambda_\parallel}$, and also assume that $\eta$ is not too large,
then to leading order, we obtain:
\begin{multline}
\delta\rho_m(\mathbf{x})\approx \eta^{2}\frac{e^{-\eta |\mathbf{x}|}}{
\sqrt{\eta|\mathbf{x}|}}\left (\delta_{m,0} -\frac{s}{2\lambda_\parallel} e^{-\frac{|m|s}{\lambda_\parallel}}\right )
\\ = \eta^{2}\frac{e^{-\eta |\mathbf{x}|}}{
\sqrt{\eta|\mathbf{x}|}}\alpha_m 
\label{eq:fluct}
\end{multline}
At large distances, the in- (out-of) plane fluctuation is positive (negative) which is 
expected by neutrality since a positive charge at the origin will attract 
negative (positive) charges towards the origin in the same (different) 
layers.  The in--plane density fluctuation has essentially the same 
Yukawa form as the single layer problem.  The screening
length is slightly renormalized from the in--plane value of
$\kappa=\sqrt{\frac{4\pi\rho_{0}}{T}}$ to 
$\eta\approx\kappa(1-\frac{s}{4\lambda})$.  However, a more striking
difference is that if we form an infinite stack by placing a test vortex at
the origin of \emph{every} layer, then the sum rule indicates that the 
stack will be completely screened while in the case of decoupled layers,
a each test charge will influence a density fluctuation in its own layer
of the 2d Yukawa form.  In this sense, the high temperature phase of the
infinite layer model does not correspond to a complete layer decoupling but 
retains some of its three--dimensional features.

\section{Conclusion}

In conclusion, we have shown that the superconductor-to-normal phase
transition in an infinite--layered, type--II superconductor, in the absence of Josephson
coupling but in the presence of electromagnetic coupling, is a Kosterlitz--Thouless
transition.  The jump in the in--plane superfluid stiffness, which is a universal
quantity in the single layer problem, acquires a small non--universal correction.  
We find that the phase transition is driven by the unbinding of two--dimensional pancake
vortices but both the low and high temperature phases show three--dimensional
characteristics.  

A natural topic for future work is to find more connections with experiment,
including ways to distinguish the electromagnetically coupled problem from the
single layer case.  A more pressing issue would be to explore how these conclusions
are affected by having a small but nonzero Josephson coupling and/or other
mechanisms of coupling the layers.  

\section{Acknowledgements}

This work was initiated while one of us (KSR) was a postdoctoral 
researcher at the University of Illinois at Urbana--Champaign in the
group of Prof. Eduardo Fradkin.  It is a pleasure to acknowledge
Vivek Aji, Eduardo Fradkin, David Huse, and Gil Refael for useful 
discussions.  This work has been supported by
the National Science Foundation through the grants 
DMR-0213706, DMR-0748925, and DMR-0758462.

\appendix

\section{Relation of Ginzburg-Landau and Coulomb gas models}  
\label{app:GLtoCoulomb}

In this Section, we review how the Coulomb gas description of a layered superconductor
arises from a phenomenological free energy functional of the Ginzburg-Landau type.  
Variants of this derivation may be found in a number of references, including the original
paper of Efetov\cite{Efetov79}.  Our presentation closely follows Ref.~\onlinecite{Fischer91}.

Our starting point is the Lawrence-Doniach model of layered superconductors\cite{Lawrence71} where the system is modeled as a discrete set of superconducting layers stacked in the $z$ direction.  
The layers are assumed to have the same thickness $d$ and are uniformly spaced with interlayer separation $s$.  With each layer $n$, we associate a superconducting order parameter $\Psi_n = |\Psi_n| e^{i\theta_n}$ which we assume does not vary in the $z$ direction within a layer.  The
Lawrence-Doniach free energy functional is then given by: 
\begin{widetext}
\bea
\mathcal{F}(\Psi_n,\mathbf{A})&=& d \sum_n \int d^3r \delta(z-z_n) \Bigl[\alpha |\Psi_n|^2 + \frac{\beta}{2} |\Psi_n|^4 + \frac{1}{2m_{\parallel}^*}\Big |\Bigl(-i\hbar\nabla_{\parallel} - \frac{e^*}{c}\mathbf{A_{\parallel}}\Big)\Psi_n\Big |^2 \nonumber \\ &+& \frac{\hbar^2}{2m_z^* s^2}\Big |\Psi_{n+1}\exp\Bigl(-i\frac{e^*}{c}\int_{ns}^{(n+1)s} dz A_z\Bigr) - \Psi_n \Big |^2\Bigr] + \frac{1}{8\pi}\int d^3r (\nabla\times\mathbf{A})^2,
\label{eq:LD}
\eea
\end{widetext}
where the subscripts $\parallel$ and $z$ refer to the two in-plane and one out-of-plane coordinates respectively and $e^*=2e$ and $m_{\parallel,z}^*=2m_{\parallel,z}$ are respectively the charge and effective masses of a Cooper pair.  The difference between Eq.~(\ref{eq:LD}) and the usual Ginzburg-Landau functional is that the order parameter fields are only defined within the layers so that the kinetic energy associated with the $z$ direction is discretized.  Note, however, that the magnetic field is defined everywhere in space.    This approach differs from anisotropic Ginzburg-Landau theory, where the order parameter is defined everywhere and the $z$ direction still has a continuum description.  We expect Eq.~(\ref{eq:LD}) to accurately describe highly anisotropic superconductors, such as the high-$T_c$ compounds, but we are not aware of a precise way in which to derive Eq.~(\ref{eq:LD}) as the limit of an anisotropic Ginzburg-Landau model.

Next, we assume the amplitude of the order parameter is constant in each layer, i.e.\ $|\Psi_n|^2=n_s^{*}=\frac{n_s}{2}$, where $n_s$ is the number of superconducting electrons per unit volume in a layer.  This assumption clearly breaks down within the core of a vortex, which is a region with a radius of order $\xi_{\parallel}\equiv\frac{\hbar^2}{2m_{\parallel}^{*}|\alpha|}$, the in-plane coherence length.  For the type II superconductors of interest in the present work, $\xi_{\parallel}$ is small compared to the in-plane magnetic penetration depth $\lambda_{\parallel} \equiv (\frac{m_\parallel c^2}{4\pi \langle n_s \rangle e^2})^{1/2}$, which is the other in-plane length scale of interest; here $\langle n_s \rangle = n_s \frac{d}{s}$ is the average number density of superconducting electrons over the whole sample volume.\cite{Clem91}  Therefore, if the smallest length scale we are interested in is of order $\lambda_{\parallel}$ and if we further assume that the concentration of vortices is dilute, it seems reasonable to neglect amplitude fluctuations. 

The term in Eq.~(\ref{eq:LD}) involving $m_z$ may then be viewed as a Josephson coupling between the phase variables in adjacent layers.  For the highly anisotropic materials which motivate the present work, $m_z\gg m_{\parallel}$ so we expect the Josephson coupling to be very small.  In this paper, we assume the Josephson coupling is identically zero and hence ignore this term.  

With these simplifications, Eq.~(\ref{eq:LD}) leads to the following effective action:
\begin{widetext}
\bea
\mathcal{F}(\theta_n, \mathbf{A})&=& \frac{\rho_s}{2}\int d^3r \sum_n \delta(z-ns) (\nabla_\parallel\theta_n - \frac{2\pi}{\phi_0} \mathbf{A_\parallel})^2 + \frac{1}{8\pi}\int d^3r (\nabla\times \mathbf{A})^2 
\label{eq:F}
\eea
\end{widetext}
where $\theta_n$ is the phase of the order parameter in layer $n$; $\rho_s=\frac{\hbar^2 n_s^{*} d}{m_\parallel^{*}}$ is the 2D superfluid stiffness of a layer;  and $\phi_0 = \frac{hc}{2e}$.  The interaction between layers is implicit in the second term.  

The next step is to determine the $\mathbf{A}$ which minimizes the functional (\ref{eq:F}).  Once this is obtained, we can rewrite Eq.~(\ref{eq:F}) solely in terms of the order parameter.  Taking the functional derivatives and imposing Coulomb gauge ($\nabla\cdot \mathbf{A}=0$) gives the following equations:
\bea
\nabla^2 A_z &=& 0 \label{eq;eomz}\\
\nabla^2 \mathbf{A_\parallel} &=& \frac{1}{\Lambda}\sum_n \delta(z-ns)(\mathbf{A_\parallel}-\frac{\phi_0}{2\pi}\nabla_\parallel \theta_n)\nonumber\\ \label{eq:eomparallel}
\eea
where $\frac{1}{\Lambda}=4\pi \rho_s (\frac{2\pi}{\phi_0})^2=\frac{s}{\lambda_{\parallel}^2}$.  Taking the Fourier transform of Eq.~(\ref{eq:eomparallel}) gives:
\bea
\mathbf{A_\parallel}(\mathbf{q},k)=-\frac{\mathbf{\alpha_\parallel}(\mathbf{q},k)-\mathbf{\varphi_\parallel}(\mathbf{q},k)}{\Lambda (q^2+k^2)}
\label{eq:FTeomparallel}
\eea
where $\mathbf{q}$ and $k$ are the momenta conjugate to the in-plane ($\mathbf{x_\parallel}$)
and out-of-plane ($z$) coordinates and:
\bea
\mathbf{\alpha_\parallel}(\mathbf{q},k)&\equiv&\sum_n e^{-ikns}\mathbf{A_{\parallel}}(\mathbf{q},ns)=\frac{1}{s}\sum_m \mathbf{A_{\parallel}}(\mathbf{q},k+\frac{2\pi m}{s})\nonumber\\ \label{eq:alpha}\\
\mathbf{\varphi_\parallel}(\mathbf{q},k)&\equiv&\sum_n e^{-ikns}\frac{\phi_0}{2\pi}\nabla_\parallel\theta_n(\mathbf{q})
\label{eq:varphi}
\eea
where $\mathbf{A_{\parallel}}(\mathbf{q},ns)$ means that the Fourier transform is only in the in-plane direction.  We can write the analog of Eq.~(\ref{eq:FTeomparallel}) with $k$ replaced by $k+(2\pi m)/s$, where $m$ is an integer.  Notice that $\mathbf{\alpha_\parallel}(\mathbf{q},k+(2\pi m)/s)=\mathbf{\alpha_\parallel}(\mathbf{q},k)$ and $\mathbf{\varphi_\parallel}(\mathbf{q},k+(2\pi m)/s) = \mathbf{\varphi_\parallel}(\mathbf{q},k)$.  Therefore, summing both sides over $m$, and using Eq.~(\ref{eq:alpha}), we obtain the minimizing $\mathbf{A}$:
\bea
\mathbf{A_\parallel}(\mathbf{q},k) &=&\frac{\mathbf{\varphi_\parallel}(\mathbf{q},k) }{(q^2+k^2)(\Lambda+L(\mathbf{q},k))}\nonumber\\ \label{eq:aparallel}\\
A_z(\mathbf{q},k) &=&  -\frac{\mathbf{q}\cdot\mathbf{\varphi_\parallel}(\mathbf{q},k)}{k(q^2+k^2)(\Lambda+L(\mathbf{q},k))} \nonumber \\ \label{eq:az}
\eea
where the last expression follows from the gauge constraint and
\bea
L(\mathbf{q},k)&\equiv&\frac{1}{s}\sum_m \frac{1}{q^2+(k+\frac{2\pi m}{s})^2}
\nonumber\\ &=& \frac{1}{2q} \frac{\sinh qs}{\cos ks - \cosh qs} 
\label{eq:L}
\eea
where the last identity is obtained by (standard) complex analysis methods.  Substituting Eqs. ~(\ref{eq:aparallel}) and (\ref{eq:az}) into Eq.~(\ref{eq:F}) will give an effective action in terms of the 
order parameter itself.  After some tedious but straightforward algebra, one obtains:
\begin{widetext}
\bea
\mathcal{F} &=& \frac{\phi_0^2}{32\pi^3\Lambda}\sum_{m,n}\int \frac{d^2q}{(2\pi)^2}\Biggl[\Biggl(\frac{\delta_{m,n}|\mathbf{q}\times\nabla_\parallel\theta_n(\mathbf{q})|^2}{q^2} - \int\frac{dk}{2\pi} \frac{[\mathbf{q}\times\nabla_\parallel\theta_m(\mathbf{q})][\mathbf{q}\times\nabla_\parallel\theta_n(\mathbf{q})] e^{ik(m-n)s}  }{q^2(q^2+k^2)(\Lambda+L(\mathbf{q},k))} \Biggr)\nonumber\\ &+& 
\Biggl(\frac{\delta_{m,n}|\mathbf{q}\cdot\nabla_\parallel\theta_n(\mathbf{q})|^2}{q^2}-\int\frac{dk}{2\pi}\frac{[\mathbf{q}\cdot\nabla_\parallel\theta_m(\mathbf{q})][\mathbf{q}\cdot\nabla_\parallel\theta_n(\mathbf{q})] e^{ik(m-n)s}  }{q^2(q^2+k^2)(\Lambda+L(\mathbf{q},k))}\Bigl(1-\frac{q^2\Lambda}{k^2(\Lambda+L(\mathbf{q},k))}\Bigr) \Biggr)\Biggr] \label{eq:swvort}
\eea
\end{widetext}
The two sets of large ``()" brackets indicate that the free energy functional decouples into two parts
which depend respectively on the divergence-free and irrotational parts of the fields $\{ \nabla_\parallel\theta_n \}$.  These correspond respectively to the ``vortex" and ``spin-wave" excitations of the order parameter field, which appear in the conventional analysis of the 2$d$XY model.\cite{Kosterlitz74}  
We make the usual assumption that the phase diagram is determined by the vortex part so will ignore the spin waves from now on.  

We make the identification $2\pi \rho_v(\mathbf{q},n)\equiv\mathbf{q}\times\nabla_\parallel\theta_n(\mathbf{q})$ where $\rho_v (\mathbf{x},n)$ is the vortex number density per unit area in layer $n$. \footnote{To see this, note that a unit strength vortex at the origin is represented by the phase field $\nabla\theta(\mathbf{x}) = \frac{\hat{\phi}}{|\mathbf{x}|}$.  For a contour $C$ encircling the singularity, $\oint_C \nabla\theta\cdot \vec{dl} = 2\pi$.  In the continuum case, this becomes $\oint_C \nabla\theta\cdot \vec{dl} = 2\pi\int\rho_v(\mathbf{x}) dA$.  However, by Green's theorem, we know that $\oint_C \nabla\theta\cdot \vec{dl} = \int (\nabla\times\nabla\theta)\cdot \vec{dA}$.  The identification $2\pi \rho_v(\mathbf{x}) = (\nabla\times\nabla\theta)\cdot \hat{z}$ follows.}  After performing the 
$k$ integration\footnote{To do the integral, it is helpful to first break the range $[-\infty,\infty]$ into separate integrals over the ranges $[-3\pi/s,\pi/s]$, $[-\pi/s, \pi/s]$, $[\pi/s,3\pi/s]$ and so on.  In the integral over $[(2a-1)\pi/s, (2a+1)\pi/s]$, make the variable change $k\rightarrow k+2\pi a/s$, so that all integrals are from $[-\pi/s,\pi/s]$.  Using Eq.~(\ref{eq:L}) and the fact that $L(\mathbf{q},k+2\pi a/s)=L(\mathbf{q},k)$, the separate integrals can be combined into one which can be performed by contour integration over the unit circle.}, the vortex part of the free energy is given by:
\beq
\mathcal{F}_v =\frac{1}{2}\sum_{m,n}\int d^2x d^2y \rho_v(\mathbf{x},m)\rho_v(\mathbf{y},n) V_{mn}(\mathbf{x-y})
\eeq       
where the vortex-vortex interaction is given by:
\beq
V_{mn}(\mathbf{x})=\frac{\phi_0^2 s}{4\pi \lambda_\parallel^2}\int \frac{d^2q}{(2\pi)^2}\frac{e^{i\mathbf{q\cdot x}}}{q^2} [\delta_{mn} - W_{mn}(\mathbf{q}) ]
\label{eq:Vmn}
\eeq
and 
\beq
W_{mn}(\mathbf{q}) = \frac{s \sinh (qs)}{2\lambda_\parallel^2 q}\frac{(G_q-(G_q^2-1)^{1/2})^{|m-n|}}{(G_q^2-1)^{1/2}}
\eeq
where $G_q = \cosh (qs) + \frac{s \sinh (qs)}{2 \lambda_\parallel^2 q}$.  While this interaction has a complicated appearance, our interest is primarily in its long distance (relative to $\lambda_\parallel$) behavior, which was obtained in Ref.~\onlinecite{Scheidl92}:
\bea
V_{mn}(|\mathbf{x}|) \simeq -\frac{\phi_0^2 s}{8\pi^2\lambda_\parallel^2}\alpha_{|m-n|} \ln {|x|} 
\label{eq:vmn}
\eea
where:
\beq
\alpha_n \simeq \delta_{n,0} - \frac{s}{2\lambda_\parallel}e^{-\frac{s}{\lambda_\parallel}|n|}
\eeq
and the ``$\simeq$'' sign indicates that this result is to leading order in $\frac{s}{\lambda_\parallel}$, which is assumed to be a small parameter.  For a discrete set of vortices, $\rho_v(\mathbf{x},n)=\sum_i m_i \delta(\mathbf{x-x_i})$ where $m_i=\pm 1, \pm 2,\dots$ and  $\mathbf{x_i}$ are the strength and position respectively of the $i$th vortex in layer $n$.  If, in addition to the interaction energy in Eq.~(\ref{eq:vmn}), we also assume that each vortex has a self-energy associated with its core, which we may represent through a fugacity, we arrive at the partition function given by Eq.~(\ref{eq:Z1}).   

\section{Relation of Coulomb gas and sine Gordon models}
\label{app:duality}

The generalized Coulomb gas model discussed in the text may be formulated as a sine Gordon field theory, as in the two--dimensional case.\cite{Ohta79}:
\begin{multline}
S[\{z_\mathbf{n}\},\boldsymbol{\phi}] = -\sum_{i,j}\frac{g_{ij}}{2}\int d^2x \nabla\phi_i\cdot\nabla\phi_j \\ + \sum_\mathbf{n}
\frac{z_\mathbf{n}}{\tau^2}\int d^2x \cos[\mathbf{n}\cdot\boldsymbol{\phi}(\mathbf{x})].
\label{eq:action}
\end{multline}
In this Appendix, we establish the equivalence of this expression with Eqs.~(\ref{eq:Z3}) and (\ref{eq:Z4}).  In Appendix \ref{app:flow}, we analyze the phase diagram of this action using
the renormalization group.  Here $i,j$ are layer indices and $\tau$ is a short distance cutoff of order the in--plane penetration depth $\lambda_\parallel$.  The matrix $\mathbf{g}$ is defined so its inverse $\mathbf{g^{-1}}$ is related to the coupling matrix for the layered Coulomb gas, given in Eq.~(\ref{eq:interactions}): $g^{-1}_{ij}/2\pi=\beta q^2\alpha_{|i-j|}$.  The factor 
$z_\mathbf{n}=2y_\mathbf{n}$, where $y_\mathbf{n}$ is the fugacity of an extended object indexed by the occupation vector $\mathbf{n}$ as discussed in the text.  The vector
$\boldsymbol{\phi}=(\dots,\phi_1,\phi_2,\dots)$ where $\phi_i$ is a sine--Gordon field 
corresponding to layer i.

We begin by writing Eq.~(\ref{eq:action}) as a partition function $\mathcal{Z}=\int D\boldsymbol{\phi} e^{S[\{z_\mathbf{n}\},\boldsymbol{\phi}]}$ and expanding the cosine terms:
\begin{multline}
\mathcal{Z}=\mathcal{Z}_0 \Bigl\langle \prod_\mathbf{n} \Bigl [1+\frac{z_\mathbf{n}}{\tau^2}\int d^2x \cos(\mathbf{n}\cdot\boldsymbol{\phi}(\mathbf{x}))\\  + \frac{z_\mathbf{n}^2}{2! \tau^4}\int\int d^2x d^2y \cos(\mathbf{n}\cdot\boldsymbol{\phi}(\mathbf{x}))\cos(\mathbf{n}\cdot\boldsymbol{\phi}(\mathbf{y})) +\dots\Bigr] \Bigr\rangle_{0}.
\label{eq:actionexpand}
\end{multline} 
Here $\mathcal{Z}_0=\int D\boldsymbol{\phi} e^{S_0[\boldsymbol{\phi}]}$ where $S_0[\boldsymbol{\phi}]=-\sum_{i,j}\frac{g_{ij}}{2}\int\nabla\phi_i\cdot\nabla\phi_j$ and $\langle\dots\rangle_0$ denotes an average over the \emph{full}
Gaussian measure.  The Coulomb gas partition function follows from calculating these averages.  We begin by writing the cosine terms as:
\beq
\cos [\mathbf{n}\cdot\boldsymbol{\phi}(\mathbf{x})] = \frac{1}{2}\sum_{\mathbf{n}(\mathbf{x})=\pm \mathbf{n}} e^{i\mathbf{n}(\mathbf{x})\cdot\boldsymbol{\phi}(\mathbf{x})}. 
\eeq 
The averages involve calculating expressions like $\langle \exp{[i\sum_{a=1}^{N} \mathbf{n}_a(\mathbf{x}_a)\cdot\boldsymbol\phi(\mathbf{x}_a)]}\rangle_0$.  Performing
the Gaussian integral:
\begin{multline}
\langle e^{[i\sum_{a=1}^{N} \mathbf{n}_a(\mathbf{x}_a)\cdot\boldsymbol\phi(\mathbf{x}_a)]}\rangle_0 \\ = \exp \Bigl (-\frac{1}{2}\sum_{kl} N_{k}N_{l}\langle\phi_k(0)\phi_l(0)\rangle \\
-\frac{1}{2}\sum_{abkl} n_{ak}(\mathbf{x}_a)n_{bl}(\mathbf{x}_b) (\langle\phi_k(\mathbf{x}_a)\phi_l(\mathbf{x}_b)\rangle-\langle\phi_k(0)\phi_l(0)\rangle)
\Bigr)
\label{eq:average}
\end{multline}  
where $k,l$ are layer indices; $a,b$ are indices denoting the $N$ objects $\{ \mathbf{n}_{a} \}$ entering the average; $N_k=\sum_a n_{ka}$ is the total vorticity in layer $k$ due to these $N$
objects; and the two-point function is:
\bea
\langle \phi_k(x)\phi_l(y) \rangle = g_{kl}^{-1} \int \frac{d^2q}{(2\pi)^2} \frac{e^{i\mathbf{q}\cdot(\mathbf{x}-\mathbf{y})}}{q^2}
\label{eq:2pt-1}
\eea
Notice that $\langle\phi_k (0)\phi_l(0)\rangle = \frac{g_{kl}^{-1}}{2\pi} \ln \frac{L}{\tau}+\dots$.
Hence the first sum in the exponential is:
\begin{equation}
-\frac{1}{2}\sum_{kl} N_{k}N_{l}\langle\phi_k (0)\phi_l(0)\rangle = -\frac{1}{2}\ln\frac{L}{\tau}\mathbf{N}^{\mathrm{T}}\mathbf{g^{-1}}\mathbf{N} +\dots
\end{equation}
where ``$\dots$" are terms subleading in $L$.
The form of the interaction (Eq.~(\ref{eq:interactions})) and the sum rule (Eq.~(\ref{eq:sumrule}))
ensure that the coupling matrix $\mathbf{g^{-1}}$ is positive definite.\footnote{To see this, note that one test of whether a symmetric matrix is positive definite is if (a) its diagonal entries are positive and (b) the diagonal entry is larger than the sum of the absolute values of the other entries in its row (or column).  A second test is that a symmetric matrix is positive definite if and only if all of the leading principal minors of the matrix are positive (i.e. which means the following matrices have positive determinants: the upper left 1-by-1 column, the upper left 2-by-2 column, and so on up to the matrix itself).  The sum rule, and the fact that the magnitudes of the couplings monotonically decrease with interlayer distance, ensures, by the first test, that for a finite number of layers, the matrix $\mathbf{g^{-1}}$ is positive definite.  As these are the leading minors of the infinite matrix, the second test ensures that the positive--definiteness holds in the infinite layer
case as well.}  Therefore, $\mathbf{N}^{\mathrm{T}}\mathbf{g^{-1}}\mathbf{N}>0$, implying the average in Eq.~(\ref{eq:average}) will be zero as $L\rightarrow\infty$ \emph{unless} $\mathbf{N}=\mathbf{0}$.  Therefore, the only configurations of objects $(\mathbf{n}_a,\cdots,\mathbf{n}_N)$
with nonzero expectation value are those that satisfy vortex neutrality in \emph{each} layer,
a stronger condition than overall vortex neutrality.  If this constraint is met, then Eq.~(\ref{eq:average}) becomes:
\beq
\exp [-\frac{1}{2}\sum_{a,b,k,l} n_{ak}(\mathbf{x}_a)n_{bl}(\mathbf{x}_b) V_{kl}(\mathbf{x_a}-\mathbf{x_b})]  
\label{eq:interaction}   
\eeq
where $k$ and $l$ indicate the planes where pancakes with strengths $n_{ak}$ and $n_{bl}$
and in--plane coordinates $\mathbf{x}_a$ and $\mathbf{x}_b$ reside and:
\bea
V_{kl}(\mathbf{x_a}-\mathbf{x_b}) &=& \langle \phi_k(\mathbf{x_a})\phi_l(\mathbf{x_b}) \rangle - \langle\phi_k(0)\phi_l(0)\rangle
\nonumber \\ &=& -\frac{g_{kl}^{-1}}{2\pi} \ln \frac{|\mathbf{x_a}-\mathbf{x_b}|}{\tau} + \dots
\eea
where ``..." are terms that are subleading for large $|\mathbf{x_i}-\mathbf{x_j}|$.   

The term we are considering will also have a prefactor.  Suppose the $N$ objects entering
the average in Eq.~(\ref{eq:average}) include $N_{\mathbf{n}_a}$ objects of type $\mathbf{n}_a$
for $a=1,\dots,N$.  There will then be a factor of $\prod_{a=1}^N [z_{\mathbf{n}_a}^{N_{\mathbf{n}_a}}/(N_{\mathbf{n}_a})!]$ from the Taylor expansion and a factor
of $1/2^N$ from writing the $N$ cosines as exponentials.  Eq.~(\ref{eq:average}) is the integrand of a $2N$ dimensional spatial integral, which may be viewed as summing over different spatial configurations of these $N$ objects.  If we choose to write this as a sum 
over \emph{indistinguishable} configurations, then there will also be a factor $\prod_{a=1}^N (N_{\mathbf{n}_a})!$ for the number of configurations (contained in the integral) identical to having the objects $\{\mathbf{n}_a\}$ at spatial positions $\{\mathbf{x}_a\}$.  Combining everything (and dropping
the unimportant factor $\mathcal{Z}_0$): we may rewrite Eq.~(\ref{eq:action}):  
\beq
\mathcal{Z}=\sum_{\{\mathbf{n}\}}\prod_{\mathbf{n}}
(z_{\mathbf{n}}/2)^{N_\mathbf{n}} 
\sum_{\{c\}}\exp (-\sum_{i\neq j} V_{\mathbf{n}_i,\mathbf{n}_j})
\label{eq:part-SG}
\eeq
where the first sum is over sets of objects $\{ \mathbf{n} \}$ satisfying vortex neutrality in
each layer and the second is over indistinguishable spatial configurations $\{ c\}$ of
these objects.  Comparing Eqs.~(\ref{eq:interaction})--(\ref{eq:part-SG}) with Eq.~(\ref{eq:Z3}), we see that the sine Gordon theory is equivalent to our generalized Coulomb gas if we make the following identifications:
\beq
y_{\mathbf{n}} = z_{\mathbf{n}}/2
\label{eq:ytoz}
\eeq
\beq
\beta q^2\alpha_{|i-j|} = g_{ij}^{-1}/2\pi
\label{eq:alphatog}
\eeq
as asserted.  Using Eq.~(\ref{eq:core2}) to expand the fugacities, we see that the sine--Gordon model of Eq.~(\ref{eq:action}) is also identical to the partition function in Eq.~(\ref{eq:Z4}).

The final point to note is that Eq.~(\ref{eq:interaction}) takes the logarithmic form at distances
large compared to $\tau$ but will vanish as $|\mathbf{x}_a-\mathbf{x}_b|\rightarrow 0$.  In
contrast, the interaction that enters Eqs.~(\ref{eq:Z3}) and (\ref{eq:Z4}) is a hard--core 
interaction (the particles are assumed to be at least distance $\tau$ laterally apart).  As 
discussed in Ref.~[\onlinecite{Knops80}], this is not actually a problem as a sine--Gordon
model literally equivalent to a hard--core Coulomb gas is possible with a slight renormalization
of the coupling constants $g_{ij}$ that will not affect our calculations (since these corrections will manifest 
as higher order terms in the RG analysis).    

\section{Derivation of flow equations}
\label{app:flow}

In this Section, we derive the RG flow equations presented in the main text.  Our starting point is the action for the layered sine Gordon model  (Eq.~(\ref{eq:action})) which, as shown in Appendix \ref{app:duality},  is equivalent to the layered Coulomb gas (Eq.~(\ref{eq:Z3})) discussed in the text.  We analyze this action using an extension of the momentum shell renormalizaton group approach discussed in  Ref.~\onlinecite{Knops80}.  The calculation applies in the small fugacity
limit.

The first step is to write the action (\ref{eq:action}) as $S[\{ z_\mathbf{n}\} ,\boldsymbol{\phi}]=S_0[\boldsymbol{\phi}]+S_1[\{z_\mathbf{n}\},\boldsymbol{\phi}]$, where $S_0$ is the Gaussian term.  The fields $\{ \phi_i \}$ are written as a sum of fast and slow modes, i.e.  $\phi_i=\phi_{i,<} + \phi_{i,>}$ where:
\beq
\phi_{i,<} (\mathbf{x}) = \int _{|\mathbf{q}|\in [0,\frac{\Lambda}{s}]} \frac{d^2q}{(2\pi)^2} e^{i\mathbf{q}\cdot\mathbf{x}}\phi_i(\mathbf{q})
\eeq
\beq
\phi_{i,>} (\mathbf{x}) = \int _{|\mathbf{q}|\in [\frac{\Lambda}{s},\Lambda]} \frac{d^2q}{(2\pi)^2} e^{i\mathbf{q}\cdot\mathbf{x}}\phi_i(\mathbf{q})
\eeq
Here $\phi_i(\mathbf{q})$ is the Fourier transform of $\phi_i(\mathbf{x})$, $\Lambda \sim \frac{1}{\tau}$ is an ultraviolet cutoff, and $s=1+\epsilon$ is a rescaling parameter.  The idea is to integrate over the fast modes to get an effective action for $\boldsymbol{\phi}_< (\mathbf{x})$.  The Gaussian term separates to give: $\mathcal{Z}=\int \mathcal{D}\boldsymbol{\phi}_< e^{S_0[\boldsymbol{\phi}_<]}\int \mathcal{D}\boldsymbol{\phi}_> e^{S_0[\boldsymbol{\phi}_>]} e^{S_1[\{z_\mathbf{n}\},\boldsymbol{\phi}_<,\boldsymbol{\phi}_>]}=A\int \mathcal{D}\boldsymbol{\phi}_< e^{S_0[\boldsymbol{\phi}_<]}\langle e^{S_1[\{z_\mathbf{n}\},\boldsymbol{\phi}_<,\boldsymbol{\phi}_>]}\rangle_>$ where the constant $A=\int \mathcal{D}\boldsymbol{\phi}_> e^{S_0[\boldsymbol{\phi}_>]}$ will be dropped because it does not affect the critical properties of the model.  The subscript ($>$) on the average, which we will also drop, denotes that the Gaussian average is over the fast modes only.  

We can write the average as $\langle e^{S_1[\{z_\mathbf{n}\}, \boldsymbol{\phi}]}\rangle=e^{S'_1[\{z'_\mathbf{n}\}, \boldsymbol{\phi}_<]}$ where the relation between $S_1$ and $S'_1$ may be expressed as a cumulant expansion:
\begin{multline}
S'_1 = \langle S_1\rangle + \frac{1}{2}(\langle S_1^2\rangle - \langle S_1\rangle^2) +\dots \\
= \sum_\mathbf{n} \frac{z_\mathbf{n}}{\tau^2}\int d^2x \langle \cos (\mathbf{n}\cdot\boldsymbol{\phi}(\mathbf{x}))\rangle + \sum_{\mathbf{m},\mathbf{n}} \frac{z_\mathbf{m}z_\mathbf{n}}{2\tau^4} \\ \int\int d^2x d^2y \Bigl [\langle \cos (\mathbf{m}\cdot\boldsymbol{\phi} (\mathbf{x})) \cos (\mathbf{n}\cdot\boldsymbol{\phi}(\mathbf{y}))\rangle \\ - \langle \cos (\mathbf{m}\cdot\boldsymbol{\phi} (\mathbf{x})) \rangle \langle \cos (\mathbf{n}\cdot\boldsymbol{\phi} (\mathbf{y}))\rangle\Bigr] \\ + \dots.
\end{multline}
A convenient way to obtain these averages is with the useful fact:
\begin{multline}
\langle e^{\int d^2x\mathbf{J}(\mathbf{x})\cdot\mathbf{\phi}(\mathbf{x})}\rangle = 
e^{\int d^2x\mathbf{J}(\mathbf{x})\cdot\mathbf{\phi}_<(\mathbf{x})}\\
\times \exp \Bigl[\sum_{k,l}\frac{1}{2}\int\int d^2x d^2y J_k(\mathbf{x})\langle \phi_{k,>}(\mathbf{x})\phi_{l,>}(\mathbf{y})\rangle J_l(\mathbf{y}) \Bigr]
\end{multline}
where the two--point function is:
\bea
\langle \phi_{k,>}(\mathbf{x})\phi_{l,>}(\mathbf{y}) \rangle = g_{kl}^{-1} \int _{|\mathbf{q}|\in [\frac{\Lambda}{s},\Lambda]} \frac{d^2q}{(2\pi)^2} \frac{e^{i\mathbf{q}\cdot(\mathbf{x}-\mathbf{y})}}{q^2}
\label{eq:2pt}
\eea
Using these relations with $\mathbf{J}(\mathbf{x'})=i\mathbf{n}\delta(\mathbf{x'}-\mathbf{x})$, we readily obtain:
\bea
\langle \cos [\mathbf{n}\cdot\boldsymbol{\phi}(\mathbf{x})]\rangle = s^{-\frac{1}{4\pi}\sum_{kl} g_{kl}^{-1}n_kn_l}\cos [\mathbf{n}\cdot\boldsymbol{\phi}_<(\mathbf{x})]
\label{eq:linear}
\eea
and with $\mathbf{J}(\mathbf{x'})=i\mathbf{n}\delta(\mathbf{x'}-\mathbf{x})\pm i\mathbf{m}\delta(\mathbf{x'}-\mathbf{y})$:
\begin{widetext}
\begin{multline}
\langle \cos[\mathbf{m}\cdot\boldsymbol{\phi}(\mathbf{x})] \cos [\mathbf{n}\cdot\boldsymbol{\phi}(\mathbf{y})]\rangle - \langle \cos[\mathbf{m}\cdot\boldsymbol{\phi}(\mathbf{x})] \rangle\langle \cos[\mathbf{n}\cdot\boldsymbol{\phi}(\mathbf{y})] \rangle  \\ = \frac{1}{2} s^{-\frac{1}{4\pi}\sum_{kl} g_{kl}^{-1}(m_km_l+n_kn_l)}\Bigl(
[e^{-\frac{1}{2}\sum_{kl}(m_kn_l+m_ln_k)\langle\phi_{k,>}(\mathbf{x})\phi_{l,>}(\mathbf{y})\rangle}-1] \cos [\mathbf{m}\cdot\boldsymbol{\phi}_<(\mathbf{x})+ \mathbf{n}\cdot\boldsymbol{\phi}_<(\mathbf{y})] \\ + 
[e^{\frac{1}{2}\sum_{kl}(m_kn_l+m_ln_k)\langle\phi_{k,>}(\mathbf{x})\phi_{l,>}(\mathbf{y})\rangle}-1] \cos [\mathbf{m}\cdot\boldsymbol{\phi}_<(\mathbf{x})-\mathbf{n}\cdot\boldsymbol{\phi}_<(\mathbf{y})] \Bigr)\\ \approx \frac{1}{2} \Bigl(\frac{1}{2}\sum_{kl} (m_kn_l
+ m_ln_k)\langle \phi_{k,>}(\mathbf{x})\phi_{l,>}(\mathbf{y})\rangle\Bigr)\Bigl(\cos  [\mathbf{m}\cdot\boldsymbol{\phi}_<(\mathbf{x})- \mathbf{n}\cdot\boldsymbol{\phi}_<(\mathbf{y})] - \cos  [\mathbf{m}\cdot\boldsymbol{\phi}_<(\mathbf{x})+ \mathbf{n}\cdot\boldsymbol{\phi}_<(\mathbf{y})]
\Bigr)\\
\label{eq:quadratic}
\end{multline}
\end{widetext}
where in the last line, we retained terms to linear order in $\epsilon$.

As the ultraviolet cutoff $\Lambda\rightarrow\infty$, the integral (\ref{eq:2pt}) becomes arbitrarily small
except when $x\approx y$.  Therefore, we make the approximation: $\langle\phi_{i,>}(\mathbf{x})\phi_{j,>}(\mathbf{y})\rangle \approx \tau^2 \delta (\mathbf{x}-\mathbf{y}) \langle\phi_{i,>}(0)\phi_{j,>}(0)\rangle \approx \tau^2 \delta (\mathbf{x}-\mathbf{y}) \frac{g_{ij}^{-1}}{2\pi} \epsilon$ and replace the cosine operators
with the leading term in their operator product expansions.\cite{Cardy96}  For $\mathbf{m}\neq\mathbf{n}$,
this implies the replacement:
\begin{multline}
\cos  [\mathbf{m}\cdot\boldsymbol{\phi}_<(\mathbf{x})\pm \mathbf{n}\cdot\boldsymbol{\phi}_<(\mathbf{y})] 
\approx \cos  [(\mathbf{m}\pm\mathbf{n})\cdot\boldsymbol{\phi}_<(\mathbf{x})]. \\
\end{multline}
Physically, this means that two closely-spaced vortex stacks, when viewed from a distance, appear as a single stack composed of pancakes that are fusions of those in the two stacks.  When $\mathbf{m}=\mathbf{n}$, one of the terms in Eq.~(\ref{eq:quadratic}) is of this fusion type:  
\begin{multline}
\cos  [\mathbf{m}\cdot(\boldsymbol{\phi}_<(\mathbf{x})+\boldsymbol{\phi}_<(\mathbf{y}))] 
\approx \cos  [2\mathbf{m}\cdot\boldsymbol{\phi}_<(\mathbf{x})],
\end{multline}
while the other involves a nontrivial operator identification:
\begin{multline}
\cos  [\mathbf{m}\cdot(\boldsymbol{\phi}_<(\mathbf{x})-\boldsymbol{\phi}_<(\mathbf{y}))] \approx
-\frac{1}{4}\sum_{ij} m_im_j \nabla\phi_i\cdot\nabla\phi_j.
\end{multline}
Physically, this latter term means that two identical, closely-spaced vortex stacks of opposite sign, to leading order, screen one another at long distances.  However, the effect of this screening is a renormalization of
the interlayer couplings.  

Using these expressions, we may write an equation for $S'_1[\{z'_\mathbf{n}\}]$ to second order in
the fugacities:
\begin{multline}
S'_1[\{z'_\mathbf{n}\}, \boldsymbol{\phi}_<] = -\sum_{i,j}\frac{\epsilon\gamma_{ij}}{2}\int d^2x \nabla\phi_{i,<}\cdot\nabla\phi_{j,<} \\ + \sum_\mathbf{n}\frac{z'_\mathbf{n}}{\tau^2}\int d^2x \cos[\mathbf{n}\cdot\boldsymbol{\phi}_< (\mathbf{x})].
\label{eq:sprime}
\end{multline}
where:
\begin{equation}
\gamma_{ij}=\frac{1}{16\pi} \sum_{\mathbf{m}} z_{\mathbf{m}}^2 \left ( \sum_{kl} m_k m_l g_{kl}^{-1}\right ) m_i m_j
\label{eq:gamma}
\end{equation}
and
\begin{multline}
z_\mathbf{n}' = \Biggl[ z_\mathbf{n} - \frac{\epsilon}{8\pi} \sum_{\mathbf{m}} z_{\mathbf{n}-\mathbf{m}}z_\mathbf{m}\left (\sum_{kl} (n_k-m_k)m_l g_{kl}^{-1}\right)\\
+ \frac{\epsilon}{8\pi} \sum_{\mathbf{m}} z_{\mathbf{n}+\mathbf{m}}z_\mathbf{m}\left (\sum_{kl} (n_k+m_k)m_l g_{kl}^{-1}\right)\Biggr ]\\ 
\times\left(1-\frac{\epsilon}{4\pi}\sum_{kl} n_k n_l g_{kl}^{-1}\right ).
\label{eq:znprime}
\end{multline}
The final step is to restore the ultraviolet cutoff of the original problem by rescaling the length.  
The net effect of doing this, to leading order in $\epsilon$, is an additional multiplicative factor
$(1+2\epsilon)$ on the right side of Eq.~(\ref{eq:znprime}).  Combining Eq.~(\ref{eq:sprime}) with 
the Gaussian term $S_0$, we obtain an action similar to Eq.~(\ref{eq:action}) but with 
renormalized parameters.  We can use Eqs.~(\ref{eq:sprime})--(\ref{eq:znprime}) to write flow
equations for these parameters:
\begin{multline}
\frac{dz_\mathbf{n}}{d\epsilon}=\left(2-\frac{1}{4\pi}\sum_{kl}n_kn_l g_{kl}^{-1}\right)z_\mathbf{n}
\\ - \frac{1}{8\pi}\sum_\mathbf{m} z_{\mathbf{n}-\mathbf{m}}z_\mathbf{m}\left(\sum_{kl}(n_k-m_k)m_lg_{kl}^{-1}\right) \\ + \frac{1}{8\pi}\sum_\mathbf{m} z_{\mathbf{n}+\mathbf{m}}z_\mathbf{m}\left(\sum_{kl}(n_k+m_k)m_lg_{kl}^{-1}\right).
\label{eq:flowzn}
\end{multline}
and:
\begin{multline}
\frac{dg_{ij}}{d\epsilon}=\frac{1}{16\pi}\sum_\mathbf{m} z_\mathbf{m}^2\left (\sum_{kl} m_km_lg_{kl}^{-1}\right )m_im_j.
\label{eq:flowgij}
\end{multline}
It is more convenient to write the latter in terms of the inverse coupling matrix:\footnote{To obtain this
equation, it is helpful to start from the infinitesimal expression for $\mathbf{g}$:  $\mathbf{g}'=\mathbf{g}+\epsilon\boldsymbol{\gamma}=\mathbf{g}(1+\epsilon \mathbf{g^{-1}}\boldsymbol{\gamma})$, where the matrix $\boldsymbol{\gamma}$ is given by Eq.~(\ref{eq:gamma}).  Then, to leading order in $\epsilon$:  $\mathbf{g^{-1}}=(1-\epsilon \mathbf{g^{-1}}\boldsymbol{\gamma})\mathbf{g^{-1}}=\mathbf{g^{-1}}-\epsilon\mathbf{g^{-1}}\boldsymbol{\gamma}\mathbf{g^{-1}}$.}
\begin{multline}
\frac{dg_{ij}^{-1}}{d\epsilon}=\\ -\frac{1}{16\pi}\sum_\mathbf{m} z_\mathbf{m}^2\left(\sum_{kl} m_km_lg_{kl}^{-1}\right)\left(\sum_{pq} g_{ip}^{-1}m_pm_q g_{qj}^{-1}\right ).
\label{eq:flowgijprime}
\end{multline}
Eqs.~(\ref{eq:flowzn})--(\ref{eq:flowgijprime}) are natural generalizations of the usual Coulomb gas
flow equations.\cite{Nienhuis87}  However, as mentioned in the text, the fugacity $z_\mathbf{n}$ 
of an extended object $\mathbf{n}$ can be expressed in terms of the creation and interaction 
energies of the pancakes forming the stack.  This latter formulation is convenient because in this
picture, the fundamental degrees of freedom are individual pancakes, which have finite energy,
as opposed to extended objects, which have (formally) infinite energy.  The content of 
Eqs.~(\ref{eq:flowzn})--(\ref{eq:flowgijprime}) can be recast in this language via the following relation which follows from Eq.~(\ref{eq:core2}):
\begin{equation}
\frac{z_{\mathbf{n}}}{2\pi}=\prod_i y_{n_i;i} \prod_{i<j} w_{n_i,n_j;i,j} \prod_{i<j<k} w_{n_i,n_j,n_k;i,j,k} \cdots.
\label{eq:ztoyw}
\end{equation}
where $i,j,k$ are layer indices and the $y$ and $w$ variables were defined in the text.  This implies:
\begin{multline}
\frac{dz_{\mathbf{n}}}{z_{\mathbf{n}}}=\\ \sum_i \frac{dy_{n_i;i}}{y_{n_i;i}} + \sum_{i<j} \frac{dw_{n_i,n_j;i,j}}{w_{n_i,n_j;i,j}} + \sum_{i<j<k} \frac{dw_{n_i,n_j,n_k;i,j,k}}{w_{n_i,n_j,n_k;i,j,k}} \\ + \dots.
\label{eq:logderiv}
\end{multline}
The flow equation for the $y$ variables are obtained by considering Eq.~(\ref{eq:flowzn}) for the occupation vector $\mathbf{n}_{a;i}$ which has only one nonzero entry: a strength $a$ vortex in layer i.  For a translationally invariant system, the fugacity variable will not depend on our choice of layer $i$ so $y_{a; i}=y_a$ and:
\begin{multline}
\frac{dy_a}{y_a} = \frac{dz_{\mathbf{n}_{a;i}}}{z_{\mathbf{n}_{a;i}}} = 2-\frac{a^2}{4\pi}g_{00}^{-1} \\ + 
\frac{1}{8\pi}\sum_\mathbf{m} \frac{z_{\mathbf{n}_{a;i}-\mathbf{m}}z_\mathbf{m}}{z_{\mathbf{n}_{a;i}}}\left( \sum_{kl}(a\delta_{k,0}-m_k)m_lg_{kl}^{-1}\right ).
\label{eq:flowya}
\end{multline}
Using Eq.~(\ref{eq:flowya}), we can obtain the flow equations for the two--body fugacities $w_{ab;ij}$ by
considering Eq.~(\ref{eq:flowzn}) with occupation vector $\mathbf{n}_{ab;ij}$ which has only two nonzero entries corresponding to vortices of strengths $a$ and $b$ in layers $i$ and $j$.
\begin{multline}
\frac{dw_{ab;ij}}{w_{ab;ij}} = \frac{dz_{\mathbf{n}_{ab;ij}}}{z_{\mathbf{n}_{ab;ij}}} - \left (\frac{dy_a}{y_a} + \frac{dy_b}{y_b}\right) = -\left (2+\frac{ab}{2\pi} g_{ij}^{-1}\right )\\-\frac{1}{8\pi}\sum_\mathbf{m}z_\mathbf{m}
\Biggl[ \Bigl( \frac{z_{\mathbf{n}_{ab;ij}-\mathbf{m}}}{z_{\mathbf{n}_{ab;ij}}} - \frac{z_{\mathbf{n}_{ab;ij}+\mathbf{m}}}{z_{\mathbf{n}_{ab;ij}}} - \frac{z_{\mathbf{n}_{a;i}-\mathbf{m}}}{z_{\mathbf{n}_{a;i}}}+ \frac{z_{\mathbf{n}_{a;i}+\mathbf{m}}}{z_{\mathbf{n}_{a;i}}}\Bigr)\\ \times \Bigl(a\sum_l m_l g_{il}^{-1}\Bigr)
+  \Bigl( \frac{z_{\mathbf{n}_{ab;ij}-\mathbf{m}}}{z_{\mathbf{n}_{ab;ij}}} - \frac{z_{\mathbf{n}_{ab;ij}+\mathbf{m}}}{z_{\mathbf{n}_{ab;ij}}} - \frac{z_{\mathbf{n}_{b;j}-\mathbf{m}}}{z_{\mathbf{n}_{b;j}}}\\ + \frac{z_{\mathbf{n}_{b;j}+\mathbf{m}}}{z_{\mathbf{n}_{b;j}}}\Bigr)\Bigl(b\sum_l m_l g_{jl}^{-1}\Bigr)
- \Bigl(  \frac{z_{\mathbf{n}_{ab;ij}-\mathbf{m}}}{z_{\mathbf{n}_{ab;ij}}} - \frac{z_{\mathbf{n}_{ab;ij}+\mathbf{m}}}{z_{\mathbf{n}_{ab;ij}}} \\ - \frac{z_{\mathbf{n}_{a;i}-\mathbf{m}}}{z_{\mathbf{n}_{a;i}}}+ \frac{z_{\mathbf{n}_{a;i}+\mathbf{m}}}{z_{\mathbf{n}_{a;i}}} - \frac{z_{\mathbf{n}_{b;j}-\mathbf{m}}}{z_{\mathbf{n}_{b;j}}}+ \frac{z_{\mathbf{n}_{b;j}+\mathbf{m}}}{z_{\mathbf{n}_{b;j}}}\Bigr)\Bigl(\sum_{kl}
m_km_lg_{kl}^{-1}\Bigr)\\ \Biggr ].
\end{multline}
In this manner, we can systematically obtain the flow equations for each of the many--body fugacities,
though the formal expressions become increasingly complicated.  

From Eqs.~(\ref{eq:flowzn})--(\ref{eq:flowgijprime}), we see that the system has a fixed point when all of
the $z_\mathbf{n}$'s equal 0.  To probe the stability of this fixed point, we consider Eq.~(\ref{eq:flowzn}),
keeping only the linear term for the moment.  Converting to the variables
in the main text via Eqs.~(\ref{eq:ytoz})--(\ref{eq:alphatog}), we have:
\begin{equation}
\frac{dy_\mathbf{n}}{d\epsilon}=\left (2-\frac{\beta}{2}\left [q^2\sum_{kl}n_kn_l\alpha_{|k-l|}\right]\right)y_\mathbf{n}. \label{eq:flowyn}
\end{equation}
where the term in square brackets is the coefficient of the logarithmic interaction energy between two
stacks of pancake vortices characterized by the same occupation vector $\mathbf{n}$.  From Eq.~(\ref{eq:interactions}), it may be shown that this quantity is always non-negative and strictly positive for
occupation vectors $\mathbf{n}$ that are ``compact", i.e.\  whose non-zero entries all occur in a region
of finite extent.  For compact vectors, Eq.~(\ref{eq:flowyn}) indicates the corresponding fugacities will be irrelevant at zero temperature.  As the temperature is raised, the first fugacities to become marginal correspond to the vectors $\{ \mathbf{n}_{1;i} \}$, which have a unit strength pancake in one layer, the other layers being empty.  Similar considerations apply for non-compact vectors that are sparsely filled.  The magnitude of the square bracket term can, in principle, be lowered by considering non-compact vectors that are densely filled.  For example, in the case where we have a unit strength pancake in \emph{every} layer of an infinite system, the square bracket term will vanish by the sum rule.  However, in a dilute system, such
objects will not be present in the initial model.  

Therefore, in the case which concerns us, the stability of the fixed point is determined by the 
$\{ y_{\mathbf{n}_{1;i}} \}$, which via Eq.~(\ref{eq:ztoyw}) is equivalent to the single strength pancake
fugacity $y_1\equiv y$ governed by the equation:
\begin{equation}
\frac{dy}{d\epsilon} = (2-\frac{\beta q^2}{2}\alpha_0)y + O(y^3)
\label{eq:RGy}
\end{equation}
We can similarly approximate the other flow equations by keeping only the leading powers in $y$.  Because we are interested in the critical behavior, we take the ``distance to marginality" $x\equiv(\beta q^2\alpha_0 - 4)$ as an additional small parameter.  In such an expansion, Eq.~(\ref{eq:flowgijprime}) becomes:
\begin{equation}
\frac{d(\beta q^2 \alpha_n)}{d\epsilon} = -\pi y^2 \sum_p (\beta q^2 \alpha_{n-p})(\beta q^2 \alpha_p)
\label{eq:RGalpha}
\end{equation}
The simplest composite object is a pair of single strength vortices, the first in layer 1 and the second directly above in layer $m$.  The flow equation for the corresponding fugacity 
$y_{\mathbf{n}_{11;1m}}\equiv y_{1m}$ simplifies to:
\begin{equation}
\frac{dy_{1m}}{d\epsilon} = \left [ 2-\beta q^2\left(\alpha_0 + \alpha_{m-1}\right ) \right ] y_{1m} -\beta q^2 \alpha_{m-1} y^2
\label{eq:RGy1m}
\end{equation}
The only fusion term surviving the expansion is the simplest one:  single strength pancakes in layers 1 
and $m$ combining to form the object $y_{1m}$.  The equation for the two--body fugacity $w_{11;1m}\equiv w_{1m}$ becomes:
\begin{equation}
\frac{dw_{1m}}{d\epsilon} = -\left [ 2+\beta q^2 \alpha_{m-1} \right ] w_{1m} - \beta q^2 \alpha_{m-1}
\label{eq:RGw1m}
\end{equation}
In this manner, we can obtain simplified versions of the full flow equations appropriate for the physical
limit that interests us. 

There is an important technical point implicit in this procedure.  In Eq.~(\ref{eq:RGy}), we have asserted that the terms we have ignored are $O(y^3)$.  From Eq.~(\ref{eq:flowzn}), we can see that an example of such a term is $g_{11}^{-1} y_{1m} y$.  This arises from the fusion of the extended object composed of two unit strength positive pancakes in layers $1$ and $m$ with a unit strength negative pancake in layer $m$.  
Since we expect $y_{1m}\sim y^2$, this term will be of order $y^3$.  However, there are an \emph{infinite} number of such terms which, in total, could potentially overwhelm the ``leading term" whenever $y$ is
nonzero.  This differs from the usual single plane Coulomb gas where there are only a finite number of
$O(y^3)$ processes (such as, for example, the fusion of a $+2$ and $-1$ pancake as well as terms
related to the cutoff procedure\cite{Nienhuis87}). To see this is not a problem, note that in the low temperature phase, Eq.~(\ref{eq:RGy1m}) implies that $y_{1m}\sim |\alpha_m|y^2/2$.  The infinite sum converges due to the sum rule:  $\sum_{m\neq 1} y_{1m} y \sim  y^3 \sum_{m\neq 1} |\alpha_m| = \alpha_0 y^3$.  The argument is similar for higher order processes.  In the absence of such a sum rule, the pile--up
of an infinite number of higher order terms can overwhelm the leading term and hence invalidate the analysis.\footnote{This pile--up issue is sometimes relevant in theories of sliding Luttinger liquids.  E. Fradkin, private communication.}

\bibliographystyle{apsrev}
\bibliography{layer}
\end{document}